\documentclass[aps,prd,twocolumn,groupedaddress,showpacs]{revtex4-1}
\usepackage[margin=1in]{geometry}
\usepackage{graphicx, subfig}
\begin{document}
\title{Exploring the predictability of symmetric texture zeros in quark mass matrices}

\author{Rohit Verma}
\email{rohitverma@live.com}
\affiliation{Chitkara University, Himachal Pradesh 174103, India}
\date{\today}
\begin{abstract}
The predictability of  hierarchical quark mass matrices with symmetrically placed texture zeros is explored in the light of current precision measurements on quark masses and mixing data and interesting phenomenological relations among the quark mass ratios and flavor mixing angles along with CP asymmetry angle $\beta$ are deduced for different predictive textures within the framework of the Standard Model. It is shown that a single non-trivial phase in these mass matrices is sufficient enough to account for the observed CP-violation in the quark sector. In particular, it is observed that the Cabibbo angle is predominantly determined by the ratio $\sqrt{m_{d}/m_{s}}$ and corrections may also result from $\sqrt{m_{u}/m_{c}}$. Furthermore, $V_{cb}\cong \sqrt{m_d/m_b}$ or $V_{cb}\cong \sqrt{m_c/2m_t}$ along with $V_{ub}\simeq \sqrt{m_u/m_t}$ provide excellent agreements with the current mixing data and $\beta  =  -Arg\lbrace V_{td}\rbrace$ may provide a rigorous test at the B-factories for some of these texture zero structures. 
\end{abstract}

\maketitle
\section{Introduction}
Despite the remarkable success of the gauge boson sector of the Standard Model (SM) \cite{Glashow:1961tr, Weinberg:1967tq, Salam:1968rm}, its Yukawa sector continues to be poorly understood and involves a large number of parameters to explain the observed quark mass spectra, flavor mixing angles and CP-violation. All the CP violation in the Standard Model originates from the interactions of the fermions with the Higgs doublet through the hadronic part of Lagrangian given by
\begin{widetext}
\begin{equation}
-{\cal{L}}(f,H) = \sum\limits_{j,k}^3 {\left\{ {Y_{jk}^u\bar Q_{jL}^u\left( {\begin{array}{*{20}{c}}
{{\phi ^{(0)*}}}\\
{ - {\phi ^{( - )}}}
\end{array}} \right)Q_{kR}^u + Y_{jk}^d\bar Q_{jL}^d\left( {\begin{array}{*{20}{c}}
{{\phi ^{(+)}}}\\
{ {\phi ^{( 0 )}}}
\end{array}} \right)Q_{kR}^d + h.c.} \right\}} 
\end{equation}
\end{widetext}
where ${Y_{jk}^q}$ $(q=u,d)$ are referred to as Yukawa couplings and are arbitrary complex numbers. The Higgs doublet involves four real scalar fields ${\phi ^{( + )}} = {{\left( {{\phi _1} + i{\phi _2}} \right)} \mathord{\left/ {\vphantom {{\left( {{\phi _1} + i{\phi _2}} \right)} {\sqrt 2 }}} \right. \kern-\nulldelimiterspace} {\sqrt 2 }}$ and ${\phi ^{(0)}} = {{\left( {{\phi _0} + i{\phi _3}} \right)} \mathord{\left/
 {\vphantom {{\left( {{\phi _0} + i{\phi _3}} \right)} {\sqrt 2 }}} \right.
 \kern-\nulldelimiterspace} {\sqrt 2 }}$. Under the spontaneous symmetry breaking, the field ${{\phi _0}}$ is shifted to ${\phi _0} + v$, where $v$ is the vacuum expectation value of the field ${{\phi _0}}$ and the mass term of the Lagrangian becomes
\begin{widetext}
\begin{equation}
-{\cal{L}}(f,H) \to  - \sum\limits_{j,k}^3 {\left\{ {M_{jk}^u{{\bar U}_{jL}}{U_{kR}} + M_{jk}^d{{\bar D}_{jL}}{D_{kR}} + h.c.} \right\}} \left( {1 + \frac{1}{v}{\phi _0}} \right)
\end{equation}
\end{widetext}
where $M_{jk}^q =  - {{vY_{jk}^q} \mathord{\left/
 {\vphantom {{vY_{jk}^q} {\sqrt 2 }}} \right.
 \kern-\nulldelimiterspace} {\sqrt 2 }}$ are called the quark mass matrices. Unfortunately, being complex 3 $\times$ 3 structures, these matrices have a  large redundancy due to presence of 36 free parameters, as compared to ten physical observables, namely the six quark masses, three mixing angles and a CP violation phase. However, due to the absence of flavor-changing right-handed currents in the SM, it is always possible to use the polar decomposition theorem to reduce each of the general quark mass matrix from complex to product of Hermitian and a Unitary matrix, where the Uniatry matrix is reabsorbed in the redefinition of right-handed quark fields. 
 
In the Standard Model of particle physics, the W boson mass is connected with the top-quark as well as Higgs-boson masses. Clearly, a precise measurement of the mass of W boson can detect any New Physics contributions in case a deviation is observed in the measured values of $m_W$ from precision fits to SM. The recent measurements of W boson mass by ATLAS Collaboration \cite{Aaboud:2017svj} at electron-positron and proton-antiproton colliders indicate $m_W=80370\pm19~MeV$ to be well consistent with the SM predictions.

Therefore, the quark mass matrices may be considered to be Hermitian without loss of generality and the number of paramaters in these matrices can be reduced to 18 in the SM and some of its extensions, but not in its left-right symmetric extensions. However, in higher dimensional models involving Grand Unification, the fermion mass matrices are usually symmetric matrices.

The quark mass eigenvalues are subsequently obtained through the diagonalization of these mass matrices using $V_{uL}^\dag M_{u} V_{uR} = diag\lbrace {m}_{\rm u}, {m}_{\rm c}, { m}_{\rm t}\rbrace$, $V_{dL}^\dag M_{d} V_{dR} = diag\lbrace {m}_{\rm d}, {m}_{\rm s}, { m}_{\rm b}\rbrace$ and the quark flavor mixing (CKM) matrix \cite{Cabibbo:1963yz, Kobayashi:1973fv} $V$ results from the non-trivial mismatch between these diagonalizations i.e. $V=V_{uL}^\dag V_{dL}$. The task of identifying viable structures of these matrices is therefore critical since the quark mass spectra, flavor mixing angles and CP violation are all determined by these mass matrices.

In view of this, several mechanisms of fermion mass generation namely radiative mechanisms \cite{Weinberg:1972tu,Zee:1980ai}, texture zeros \cite{Weinberg:1977hb, Wilczek:1977uh,Fritzsch:1977vd,Fritzsch:1979zq, Ramond:1993kv,Ibanez:1994ig,Fritzsch:1999ee}, family symmetries \cite{Pakvasa:1977in,Froggatt:1978nt,Harari:1978yi,Yamanaka:1981pa,Wilczek:1978xi}, see-saw mechanisms \cite{Yanagida:1979as, Glashow:1979nm,GellMann:1980vs}, grand unified theories \cite{Pati:1973uk,Chen:2003zv} and extra dimensions \cite{Dienes:1998vg, ArkaniHamed:1998rs,Dvali:1999cn} have been adopted. However, in the absence of a compelling theory from "top-down" perspective, phenomenological "bottom-up" approaches have continued to play a vital role in interpreting new experimental data on fermion mixing and CP-violation.

One of the successful ansatze incorporating "texture zero" approach was initiated by Weinberg \cite{Weinberg:1977hb} and Fritzsch \cite{Fritzsch:1977vd}. A particular texture structure is said to be texture 'n' zero, if it has 'n' number of non-trivial zeros, for example, if the sum of the number of diagonal zeros and half the number of symmetrically placed off-diagonal zeros is 'n'. A texture zero in a mass matrix is a phenomenological zero that essentially represents an element in the mass matrix which is strongly suppressed as compared to other elements in the same matrix. Reasonable zeros in quark mass matrices allow us to 
establish simple and testable relations between flavor mixing angles and quark mass ratios. Therefore, a phenomenological study of possible texture zeros, and the relations so predicted, do make some sense to get useful hint about flavor dynamics responsible for quark mass generation, flavor mixing and CP-violation.

In the last few years, the precision in measurements of several vital mixing parameters associated with the CKM matrix has significantly improved and most of these are now measured with less than few percent error. In particular, the current global averages \cite{Olive:2016xmw} for best fit values of the CKM elements at 95 $\%$ CL are given by 
\begin{widetext}
\begin{equation}\label{PDG best fit}
\left| V \right| = \left( {\begin{array}{*{20}{c}}
{0.97434 \pm 0.00012}&0.22506 \pm 0.00050&0.00357 \pm 0.00015\\
{0.22492 \pm 0.00050}&{0.97351 \pm 0.00013}&0.0411 \pm 0.0013\\
{0.00875\pm 0.00033}&{0.0403\pm 0.0013}&{0.99915 \pm 0.00005}
\end{array}} \right).
\end{equation}
\end{widetext}
Likewise, the three inner angles of the unitarity triangle in $V^{}_{ud} V^*_{ub} + V^{}_{cd} V^*_{cb} + V^{}_{td} V^*_{tb} = 0$ are also quite precisely measured \cite{Olive:2016xmw}, e.g.
\begin{eqnarray}\label{inner angles}
\alpha  &\equiv& \arg \left(-\frac{V^{}_{td} V^*_{tb}}{V^{}_{ud} V^*_{ub}}\right)= {87.6^\circ}^{+3.5^\circ}_{-3.3^\circ} \; , \nonumber \\
\beta  &\equiv& \arg \left(-\frac{V^{}_{cd} V^*_{cb}}{V^{}_{td} V^*_{tb}}\right)= {21.85^\circ}\pm{0.48^\circ} \; , \nonumber \\
\gamma  &\equiv& \arg \left(-\frac{V^{}_{ud} V^*_{ub}}{V^{}_{cd} V^*_{cb}}\right)= {73.2^\circ}^{+6.3^\circ}_{-7.0^\circ} \;.
\end{eqnarray}
Based on these precise values, it has emerged \cite{Olive:2016xmw} that a single CKM phase provides a viable solution of CP violation not only in the case of K-decays but also in the context of B-decays, at least to the leading order. Furthermore, using the running quark masses at the energy scale of $M^{}_Z = 91.2~{\rm GeV}$ \cite{LEUTWYLER1996313, Xing:2011aa, Olive:2016xmw}, one obtains the following estimates for these masses with reasonable precision,
\begin{eqnarray}\label{eq:quark masses}
m_u = 1.38^{+0.42}_{-0.41}~{\rm MeV} \; ,~~
m_d = 2.82^{+0.48}_{-0.48}~{\rm MeV}\; , \nonumber \\
m_c = 0.638^{+0.043}_{-0.084}~{\rm GeV} \; , 
m_s = 57^{+18}_{-12}~{\rm MeV}\;,\nonumber \\
m_t = 172.1^{+1.2}_{-1.2}~{\rm GeV} \; ,~~ 
m_b = 2.86^{+0.16}_{-0.06}~{\rm GeV}\;, \nonumber \\
m_{u}/m_{d}=0.38-0.58,~~ m_{s}/m_{d}=17-22.\;\nonumber \\
\end{eqnarray}
Noting this significant improvement in the precision measurements of the above parameters, it becomes desirable to revisit and investigate predictive texture structures in the quark sector of the SM.

Furthermore, the quark mass spectra appear to exhibit a strong hierarchy i.e. ${ m}_{\rm u} \ll {m}_{\rm c} \ll { m}_{\rm t}$ and ${ m}_{\rm d} \ll { m}_{\rm s} \ll { m}_{\rm b}$ with the hierarchy much stronger in the 'up' sector of quarks. Likewise, the CKM matrix also satisfies a strongly hierarchical structure viz. $\vert{V_{{\rm{ub}}}}\vert<\vert{V_{{\rm{td}}}}\vert \ll
\vert{V_{{\rm{ts}}}}\vert < \vert{V_{{\rm{cb}}}}\vert\ll \vert{V_{{\rm{cd}}}}\vert < \vert{V_{{\rm{us}}}}\vert<\vert{V_{{\rm{cs}}}}\vert <\vert{V_{{\rm{ud}}}}\vert<\vert{V_{{\rm{tb}}}}\vert$, indicating that any natural description of the observed flavor mixing should translate these strong hierarchies in the quark mass spectra and the flavor mixing parameters onto the hierarchy and phase structure of the correspondence mass matrices. 

In the context of texture zeros, it was shown \cite{Branco:1988iq,Fritzsch:1997fw,Fritzsch:1999rb,Branco:1999nb,Fritzsch:2002ga} that it is always possible to have Hermitian quark mass matrices with three texture zeros which do not have any physical implications. Any additional texture zero should introduce a physical assumption and imply a testable relationship between the quark masses and the parameters of the mixing matrix. For example, six texture zeros should imply three physical relationships relating the three mixing angles with six quarks masses and the CP violation phase. Also the maximum number of such texture zeros consistent with the absence of a zero mass eigenvalue and a non degenerate quark mass spectrum is six, with three each in the "up" and "down" quark sectors. 

In particular, it was observed \cite{Desai:2000bu,Mahajan:2009wd,Ponce:2011qp,Ludl:2015lta} that all texture six zero hierarchical quark mass matrices were ruled out at the level of 3$\sigma$, which is a settled and uncontested result. Some predictive texture five zero structures were discussed in \cite{Ramond:1993kv} which were subsequently ruled out in \cite{Desai:1997hk,Kim:2004ki,Mahajan:2009wd} on hierarchy and sin$2\beta$ considerations. These structures were again reviewed in \cite{Ponce:2013nsa}, but allowed 3$\sigma$ variation of quark mixing angles and sin$2\beta$. However, $sin2\beta$ is now more precisely measured and a systematic investigation is therefore desirable. 

Some other interesting texture five zero structures were discussed in \cite{Mahajan:2009wd, Ponce:2011qp,Giraldo:2011ya,Fakay:2014rea,Giraldo:2015ffa,Giraldo:2015cpp} with conflicting claims \cite{Gupta:2011zzg,Sharma:2015jaa,Giraldo:2015dta,Sharma:2015gfa} where \cite{Gupta:2011zzg} established that all texture five zero mass matrices except Fritzsch-like texture five zero structures were not viable, a result which was contested in \cite{Giraldo:2011ya,Giraldo:2015ffa,Giraldo:2015cpp}. 

Dong-Sheng Du and Zhi Zhong Xing \cite{Du:1992iy} introduced a new structure for texture four zero quark mass matrices, which was a modification of the famous Fritzsch type texture six zero matrices \cite{Fritzsch:1977vd}. This new structure, also referred to as Fritzsch-like texture four zero was later pursued in \cite{Fritzsch:1999ee, Fritzsch:2002ga,Fritzsch:2002ga,Mei:2003jr,Xing:2003yj, Verma:2010jy,Xing:2014sja, Xing:2015sva, Fritzsch:2017tyf} and was observed to offer a promising explanation for observed flavor mixing both in the quark as well as the lepton sectors \cite{Fritzsch:1999ee, Ahuja:2007vh, Ahuja:2009jj,Verma:2010jy,Fritzsch:2011qv, Verma:2013cza,Fakay:2013gf,Verma:2014woa,Verma:2014lpa,Verma:2016qhy,Singh:2016qcf}. Initial observations on the non-viability of this structure on account of sin$2\beta$ predictions in \cite{Kim:2004ki} were investigated \cite{Verma:2009gf, Gupta:2009ur} to establish that these matrices must involve two non-trivial phases in the quark mass matrices to account for observed CP-violation in the quark sector but predicted the following relation for the Cabibbo angle  and quark mass ratios
\begin{equation}
{V_{us}} = \sqrt {\frac{{{m_d}}}{{{m_s}}}}  - \sqrt {\frac{{{m_u}}}{{{m_c}}}} {e^{i\phi }}.
\end{equation}
 
Attempts were also made \cite{Verma:2013qta} to integrate the concept of 'weak basis transformations' \cite{Branco:1988iq,Branco:1999nb} and 'naturalness' \cite{Peccei:1995fg} in quark mass matrices which were also consistent with hierarchical quark mass matrices. Some recent analyses predicted \cite{Sharma:2014tea,Gupta:2015qaw,Sharma:2015gfa,Gupta:2016hfm,Ahuja:2016khy,Ahuja:2017fhi} that texture four zero quark mass matrices in \cite{Du:1992iy} were the only viable structures in the quark sector and that the elements in these matrices were only "weakly" hierarchical with $M_{22}>>m_{2}$ \cite{Verma:2010jy} and not "strongly" hierarchical with $M_{22}\sim m_{2}$. These analyses seemed to rule out the possibility of strongly hierarchical quark mass matrices in recent context. This was recently updated in \cite{Verma:2015mgd, Verma:2016lrl} and a fine tuning of certain parameters in the quark mass matrices was observed to result in hierarchical strictures of these mass matrices.

Another, interesting numerical survey of predictive quark textures was presented recently in \cite{Ludl:2015lta} but excluded the nature of testable relations predicted by these between the quark mass ratios and mixing angles. 

In view of above developments, it becomes desirable to investigate the phenomenological implications of the current precision mixing data on predictive (Hermitian or symmetric) quark mass textures, involving the least number of phases, at least in the SM framework and the current work is an attempt in this direction. 

To this end, in the light of precision measurements, we perform a systematic phenomenological analysis of symmetrically placed texture zeros in quark mass matrices allowing 1$\sigma$ variation of mixing parameters along with heavy quark masses and upto 3$\sigma$ variation of light quark masses namely $m_{u},~ m_{d}$ and $m_{s}$ to investigate the implications of precision data for predictive texture zeros in quark sector, within the Standard Model framework. 

The 3$\sigma$ variation of light quark masses allows for an investigation of the sensitivity of certain texture zeros to these mass eigenvalues as since the latter are indirectly derived from lattice QCD calculations and may involve uncertainties from higher order perturbations \cite{Olive:2016xmw}.
\section{Standard Parametrization}
In the standard parametrization \cite{Olive:2016xmw}, $V$ can be expressed in terms of the three mixing angles $s_{ij}={sin} \theta_{ij}$, $c_{ij}={cos} \theta_{ij}$ where $\theta_{ij}$ $(i,j=1,2,3)$ lie in the first quadrant i.e. $s_{ij},c_{ij}\geq 0$ and a CP-violation phase $\delta_{13}$ associated with the flavor-changing processes in the SM, e.g.
\begin{widetext}
\begin{eqnarray}\label{choice of parametrization}
V =R_{23}R_{13}R_{12}= \left( {\begin{array}{*{20}{c}}
1&0&0\\
0&{{c_{23}}}&{{s_{23}}}\\
0&{ - {s_{23}}}&{{c_{23}}}
\end{array}} \right)\left( {\begin{array}{*{20}{c}}
{{c_{13}}}&0&{{s_{13}}{e^{  -i{\delta _{13}}}}}\\
0&{ 1}&0\\
{ - {s_{13}}{e^{i{\delta _{13}}}}}&0&{{c_{13}}}
\end{array}} \right)\left( {\begin{array}{*{20}{c}}
{{c_{12}}}&{{s_{12}}}&0\\
{ - {s_{12}}}&{{c_{12}}}&0\\
0&0&{  1}
\end{array}} \right)\; \nonumber\\
= \left( {\begin{array}{*{20}{c}}
{{c_{12}}{c_{13}}}&{{s_{12}}{c_{13}}}&{  {s_{13}}{e^{  -i{\delta _{13}}}}}\\
{-{s_{12}}{c_{23}} - {c_{12}}{s_{23}}{s_{13}}{e^{i{\delta _{13}}}}}&{ {c_{12}}{c_{23}} - {s_{12}}{s_{23}}{s_{13}}{e^{i{\delta _{13}}}}}&{  {s_{23}}{c_{13}}}\\
{ {s_{12}}{s_{23}}-{s_{13}}{c_{12}}{c_{23}}{e^{i{\delta _{13}}}}  }&{-{s_{23}}{c_{12}} - {s_{12}}{s_{13}}{c_{23}}{e^{i{\delta _{13}}}}}&{  {c_{23}}{c_{13}}}
\end{array}} \right)
= \left( {\begin{array}{*{20}{c}}
V_{ud}& V_{us}& V_{ub}\\
V_{cd}& V_{cs}& V_{cb}\\
V_{td}& V_{ts}& V_{tb}
\end{array}} \right).
\end{eqnarray}
\end{widetext}
Using $s_{13}\ll s_{23}\ll s_{12}$, at the leading order, the above structure can be reduced to the following form without losing generality \cite{Verma:2016lrl},
\begin{equation}\label{reduced CKM}
V\simeq\left( {\begin{array}{*{20}{c}}
{{c_{12}}}&{{s_{12}}}&{  {s_{13}}{e^{  -i{\delta _{13}}}}}\\
{-{s_{12}}{c_{23}}}&{{c_{12}}{c_{23}}}&{ {s_{23}}}\\
{ {s_{12}}{s_{23}}- {s_{13}}{c_{12}}{e^{i{\delta _{13}}}} }&{-{s_{23}}{c_{12}}}&{  {c_{23}}}
\end{array}} \right)
\end{equation}
and follows from $s_{12}s_{23}s_{13}\ll 1$, $s_{12}s_{13}\ll s_{23}$, $s_{23}s_{13}\ll s_{12}$, $s_{12}s_{23}\sim s_{13}$ and $c_{ij}\simeq 1$. This further allows to establish a simpler relation for the unitarity angle $\beta$ with the argument of $V_{td}$ through
\begin{eqnarray}\label{beta relation}
\beta  =  - \arg \left( {1 - \frac{{{s_{13}}{c_{12}}}}{{{s_{12}}{s_{23}}}}{e^{ i{\delta _{{\rm{13}}}}}}} \right)=-\phi,\nonumber \\
\phi=Arg\lbrace V_{td}\rbrace
\end{eqnarray}
Likewise, the Jarlskog's CP invariant parameter \cite{PhysRevLett.55.1039, Jarlskog:1985cw, Wu:1985ea} 
\begin{eqnarray}
J_{\rm{CP}}=Im[V_{us}V_{cb}V^{*}_{cs}V^{*}_{ub}]\simeq {c_{12}}s_{12}s_{23}s_{13}{\rm {sin}}\delta_{13}.\nonumber \\
\end{eqnarray}
The current global average \cite{Olive:2016xmw} for this parameter is given by ${J_{{\rm{CP}}}} = (3.06_{ - 0.20}^{ + 0.21}) \times {10^{ - 5}}$.

At the leading order, almost all of the phase information in the CKM matrix is vested in $V_{\rm {ub}}=s_{13}e^{i\delta_{13}}$ and $V_{\rm{td}}=|V_{\rm {td}}|e^{i\phi}$ and the following phase correspondences can be established 
\begin{eqnarray}\label{phase relations}
\beta = -\phi,\nonumber \\
\gamma= \delta_{13},\nonumber \\
\alpha= \pi+\phi-\delta_{13}
\end{eqnarray}
in agreement with the condition of unitarity i.e. $\alpha+\beta+\gamma=\pi$. However, these relationships may not be followed by all possibilities of texture zero quark mass matrices, e.g. Fritzsch-like texture four zeros.
\section{Hierarchical Quark Mass Matrices}
In view of the strong hierarchy of quark masses and mixing angles, we expect naturally the hierarchical structures of quark mass matrices, which in the most general case can be formulated as \cite{Verma:2015mgd}
\begin{widetext}
\begin{eqnarray}\label{eq:hierarchical}
M^{}_{\rm q} = \left(\matrix{ e^{}_{\rm q} & a^{}_{\rm q} e^{{\rm i} \alpha^{}_{\rm q}} & f^{}_{\rm q} e^{{\rm i} \gamma^{}_{\rm q}} \cr a^{}_{\rm q} e^{-{\rm i} \alpha^{}_{\rm q}} & d^{}_{\rm q} & b^{}_{\rm q} e^{{\rm i} \beta^{}_{\rm q}} \cr f^{}_{\rm q} e^{-{\rm i} \gamma^{}_{\rm q}} & b^{}_{\rm q} e^{-{\rm i} \beta^{}_{\rm q}}  & c^{}_{\rm q} \cr } \right) \sim \left(\matrix{ m^{}_1 & \sqrt{m^{}_1 m^{}_2} e^{{\rm i} \alpha^{}_{\rm q}} &\sqrt{m^{}_1 m^{}_3} e^{{\rm i} \gamma^{}_{\rm q}} \cr \sqrt{m^{}_1 m^{}_2} e^{-{\rm i} \alpha^{}_{\rm q}} & m^{}_2 & \sqrt{m^{}_2 m^{}_3} e^{{\rm i} \beta^{}_{\rm q}} \cr \sqrt{m^{}_1 m^{}_3} e^{-{\rm i} \gamma^{}_{\rm q}} & \sqrt{m^{}_2 m^{}_3} e^{-{\rm i} \beta^{}_{\rm q}}  & m^{}_3 \cr } \right),
\end{eqnarray}
\end{widetext}
where $m^{}_i$ denotes the mass of $i$-th generation up-type (or down-type) quark for ${\rm q} = {\rm u}$ (or ${\rm d}$) and each nonzero matrix element in the rightmost term in Eq.~(\ref{eq:hierarchical}) is up to a factor of ${\cal O}(1)$. 

The diagonalization of $M_{\rm q}$ can now be achieved by a unitary matrix $V^{}_{\rm q}$, which can be parametrized in terms of three rotation angles $\{\theta^{\rm q}_{12}, \theta^{\rm q}_{13}, \theta^{\rm q}_{23}\}$. We adopt a convenient parametrization of $V^{}_{\rm q} \equiv R^{}_{13}(\theta^{\rm q}_{13}) R^{}_{12}(\theta^{\rm q}_{12}) R^{}_{23}(\theta^{\rm q}_{23})$, where $R^{}_{ij}(\theta^{\rm q}_{ij})$ is the rotation matrix in the $ij$-plane with an angle $\theta^{\rm q}_{ij}$ (for $ij = 12$, $13$, $23$),  although the order of these three rotation matrices is irrelevant in the light of hierarchical quark mass matrices and rotation angles. 

Due to the hierarchical structure of $M^{}_{\rm q}$, one may now achieve the necessary correlation of the hierarchical quark flavor mixing angles $\theta_{ij}$ with the quark rotation angles $\theta^{\rm q}_{ij}$ (for $ij = 12$, $13$, $23$) associated with the quark mass ratios \cite{Verma:2015mgd}. Using $V^{\dagger}_{\rm q} M_{\rm q} V^{}_{\rm q} = \tilde{M}_{\rm q}$, where $\tilde{M}_{\rm q} = {\rm diag}\{ k_{q\rm } m^{}_1, - k_{\rm q} m^{}_2, m^{}_3\}$ is a diagonal matrix with quark mass eigenvalues and $k_{\rm q}=\pm 1$. One therefore obtains the CKM Matrix through $V_{CKM}=V^{\dagger}_{u}V_{d}$, such that
\begin{widetext}
\begin{equation}\label{CKM}
V = \left( {\begin{array}{*{20}{c}}
{c_{12}^d{e^{ - i{\phi _1}}}}&{\left( {s_{12}^d{e^{ - i{\phi _1}}} - s_{12}^u} \right)}&{\left( {s_{13}^d{e^{ - i{\phi _1}}} - s_{13}^u{e^{i{\phi _2}}} - s_{12}^u{V_{cb}}} \right)}\\
{\left( {s_{12}^u{e^{ - i{\phi _1}}} - s_{12}^d} \right)}&{c_{12}^dc_{23}^d}&{\left( {s_{23}^d - s_{23}^u{e^{i{\phi _2}}}} \right)}\\
{\left( {s_{13}^u{e^{ - i{\phi _1}}} - s_{13}^d{e^{i{\phi _2}}} - s_{12}^d{V_{ts}}} \right)}&{\left( {s_{23}^u - s_{23}^d{e^{i{\phi _2}}}} \right)}&{c_{23}^d{e^{i{\phi _2}}}}
\end{array}} \right),
\end{equation}
\end{widetext}
$\phi_{1}=\alpha_{u}-\alpha_{d}, ~~
\phi_{2}=\beta_{u}-\beta_{d}$.
From Eqs.(\ref{beta relation}-\ref{phase relations}), one obtains a general relation for the phase $\phi$ associated with $V_{td}$, i.e.
\begin{equation}
\phi  = Arg\left\{ {s_{13}^u{e^{ - i{\phi _1}}} - s_{13}^d{e^{i{\phi _2}}} - s_{12}^ds_{23}^u + s_{12}^ds_{23}^d{e^{i{\phi _2}}}} \right\}.
\end{equation}
In particular, the general expressions for the off-diagonal CKM elements are calculated as \cite{Verma:2015mgd}
\begin{eqnarray}\label{eq:approx}
V^{}_{us} &\approx& \sqrt{\frac{m^{}_d}{m^{}_s}} e^{ - {\rm i} \phi^{}_1} - \sqrt{\frac{m^{}_u}{m^{}_c}} \; , \nonumber \\
V^{}_{cd} &\approx& \sqrt{\frac{m^{}_u}{m^{}_c}} e^{ - {\rm i} \phi^{}_1} - \sqrt{\frac{m^{}_d}{m^{}_s}} \; , \nonumber \\
V^{}_{cb} &\approx& \sqrt{\frac{d^{}_{\rm d} + k_{\rm d} m^{}_s}{m^{}_b}} - \sqrt{\frac{d^{}_{\rm u} + k_{\rm u} m^{}_c}{m^{}_t}} e^{{\rm i}\phi^{}_2} \; , \nonumber \\
V^{}_{ts} &\approx& \sqrt{\frac{d^{}_{\rm u} + k_{\rm u} m^{}_c}{m^{}_t}} - \sqrt{\frac{d^{}_{\rm d} + k_{\rm d} m^{}_s}{m^{}_b}} e^{{\rm i}\phi^{}_2} \; , \nonumber \\
V^{}_{ub} &\approx& \varepsilon^{}_{\rm d} \sqrt{\frac{m^{}_d}{m^{}_b}} e^{ - {\rm i} \phi^{}_1} - \varepsilon^{}_{\rm u} \sqrt{\frac{m^{}_u}{m^{}_t}} e^{{\rm i} \phi^{}_2} - \sqrt{\frac{m^{}_u}{m^{}_c}} V^{}_{cb} \; , \nonumber \\
V^{}_{td} &\approx& \varepsilon^{}_{\rm u} \sqrt{\frac{m^{}_u}{m^{}_t}} e^{ - {\rm i} \phi^{}_1} - \varepsilon^{}_{\rm d} \sqrt{\frac{m^{}_d}{m^{}_b}} e^{{\rm i} \phi^{}_2} - \sqrt{\frac{m^{}_d}{m^{}_s}} V^{}_{ts}. \nonumber \\
\end{eqnarray}
where $\varepsilon^{}_{\rm q} \equiv f^{}_{\rm q}/\sqrt{m^{}_1 m^{}_{3}}\sim {\cal O}(1)$ in agreement with hierarchical structures in Eq.~({\ref{eq:hierarchical}) wherein $f_{\rm q}\lesssim {\cal O}\sqrt{m_1 m_3}$. 

It is clear that $s^{d}_{12}\ne 0$ is required for consistency with $\mid V_{us}\mid$ and $s^{u}_{12}$ alone is insufficient as $\sqrt{m_{u}/m_{c}}<<\mid V_{us}\mid$. For $\mid V_{cb}\mid$ one may require either $s^{u}_{23}\ne 0$ or $s^{d}_{23}\ne 0$ or both. Likewise for $\mid V_{ub}\mid$, one observes that $s^{u}_{13}\ne 0$ or $s^{d}_{13}\ne 0$ or both may contribute. 

Furthermore, since there is only one physical phase $\delta_{13}$ in the CKM matrix, it is desirable to translate this single phase onto the corresponding mass matrix and invoke a minimal phase structure for the same. As a result, we will introduce this phase in the quark rotation matrix associated with one rotation only and treat other rotation matrices as real and symmetric. In this context, it may be emphasized that the quark mass matrices are 'weak' basis dependent due to the freedom of weak basis transformations \cite{Branco:1988iq}. The diagonal basis of $M_{u}$ has already been discussed in \cite{Verma:2016lrl} and study reveals that the diagonal basis of $M_{d}$ does does not lead to any predictions for the 'up' sector \cite{Ludl:2015lta}. Quark mass matrix structures pertaining to texture zeros at the (33) position in $M_q$ are not addressed in this paper for inconsistency with the notion of hierarchical quark mass matrices.
\section{Predictive structures}
The predictive texture possibilities for hierarchical quark mass matrices that emerge in the non-diagonal basis of $M_{q}$ are listed in Table-I and their corresponding texture structures are listed in Table-II. These are further addressed below with relevant details.
\begin{table}
\begin{tabular}{c c c c c c c}
\hline
Case & $s_{13}$ & $s_{13}$ & $s_{12}$ &$s_{12}$ & $s_{23}$& $s_{23}$ \\
\hline
~& LO & NLO & LO & NLO &LO & NLO\\
\hline
I & $s^{u}_{13}$ & $(s^{d}_{12}s^{d}_{23})$ & $s^{d}_{12}$ &$\times$& $s^{d}_{23}$& $\times$ \\
\hline
II & $s^{d}_{13}$&$\times$ & $s^{d}_{12}$ &$\times$& $s^{u}_{23}$&$\times$ \\
\hline
III & $s^{u}_{13}$ &$\times$& $s^{u}_{12}$&$\times$ & $s^{d}_{23}$&$\times$ \\
\hline
IV & $s^{u}_{13}$ &$\times$& $s^{d}_{12}$ & $s^{u}_{12}$& $s^{u}_{23}$ & $\times$ \\
\hline
V & $s^{d}_{13}$ &$\times$& $s^{u}_{12}$ &$s^{d}_{12}$& $s^{d}_{23}$ &$\times$ \\
\hline
VI & $s^{d}_{13}$ &$\times$& $\times$&$s^{d}_{12}$ & $s^{d}_{23}$ & $s^{u}_{23}$ \\
\hline
VII & $\times$ &$(s^{d}_{12}s^{d}_{23})$& $s^{d}_{12},s^{u}_{12}$&$\times$ & $s^{d}_{23}$ & $\times$ \\
\hline
VIII & $\times$ &$(s^{u}_{12}s^{u}_{23})$& $s^{u}_{12},s^{d}_{12}$&$\times$ & $s^{u}_{23}$ & $\times$ \\
\hline
\end{tabular}
\caption{Trivial possibilities of quark rotation angles contributing to flavor mixing angles. LO$\equiv$ Leading Order and NLO$\equiv$ Next to Leading Order.}
\end{table}
\begin{table}
\begin{tabular}{c c c }
\hline
Case & $M_{u}$ & $M_{d}$\\ [0.5ex]
\hline
I & $\left( {\begin{array}{*{20}{c}}
{\bf{0}}&{\bf{0}}& \times \\
{\bf{0}}& \times &{\bf{0}}\\
 \times &{\bf{0}}& \times 
\end{array}} \right)$ & $\left( {\begin{array}{*{20}{c}}
{\bf{0}}& \times &{\bf{0}}\\
 \times & \times & \times \\
{\bf{0}}& \times & \times 
\end{array}} \right)$ \\
\hline
II & $\left( {\begin{array}{*{20}{c}}
 \times &{\bf{0}}&{\bf{0}}\\
{\bf{0}}& \times & \times \\
{\bf{0}}& \times & \times 
\end{array}} \right)$  & $\left( {\begin{array}{*{20}{c}}
{\bf{0}}& \times & \times \\
 \times & \times &{\bf{0}}\\
 \times &{\bf{0}}& \times 
\end{array}} \right)$  \\
\hline
III & $\left( {\begin{array}{*{20}{c}}
{\bf{0}}& \times & \times \\
 \times & \times &{\bf{0}}\\
 \times &{\bf{0}}& \times 
\end{array}} \right)$ & $\left( {\begin{array}{*{20}{c}}
 \times &{\bf{0}}&{\bf{0}}\\
{\bf{0}}& \times & \times \\
{\bf{0}}& \times & \times 
\end{array}} \right)$  \\
\hline
IV & $\left( {\begin{array}{*{20}{c}}
{\bf{0}}&{\bf{0}}& \times \\
{\bf{0}}& \times & \times \\
 \times & \times & \times 
\end{array}} \right)$ & $\left( {\begin{array}{*{20}{c}}
{\bf{0}}& \times &{\bf{0}}\\
 \times & \times &{\bf{0}}\\
{\bf{0}}&{\bf{0}}& \times 
\end{array}} \right)$ \\
\hline
V & $\left( {\begin{array}{*{20}{c}}
{\bf{0}}& \times &{\bf{0}}\\
 \times & \times &{\bf{0}}\\
{\bf{0}}&{\bf{0}}& \times 
\end{array}} \right)$ & $\left( {\begin{array}{*{20}{c}}
{\bf{0}}&{\bf{0}}& \times \\
{\bf{0}}& \times & \times \\
 \times & \times & \times 
\end{array}} \right)$ \\
\hline
VI & $\left( {\begin{array}{*{20}{c}}
 \times &{\bf{0}}&{\bf{0}}\\
{\bf{0}}&{\bf{0}}& \times \\
{\bf{0}}& \times & \times 
\end{array}} \right)$ & $\left( {\begin{array}{*{20}{c}}
{\bf{0}}&{\bf{0}}& \times \\
{\bf{0}}& \times & \times \\
 \times & \times & \times 
\end{array}} \right)$ \\
\hline
VII & $\left( {\begin{array}{*{20}{c}}
\times &\times&{\bf{0}}\\
\times&\times& {\bf{0}} \\
{\bf{0}}& {\bf{0}} & \times 
\end{array}} \right)$ & $\left( {\begin{array}{*{20}{c}}
{\bf{0}}&\times& {\bf{0}} \\
\times& \times & \times \\
{\bf{0}} & \times & \times 
\end{array}} \right)$ \\
\hline
VIII & $\left( {\begin{array}{*{20}{c}}
\times&\times& {\bf{0}} \\
\times& \times & \times \\
{\bf{0}} & \times & \times 
\end{array}} \right)$ & $\left( {\begin{array}{*{20}{c}}
{\bf{0}} &\times&{\bf{0}}\\
\times&\times& {\bf{0}} \\
{\bf{0}}& {\bf{0}} & \times 
\end{array}} \right)$  \\
\hline
\end{tabular}
\caption{Predictive texture zero possibilities.}
\end{table}
\subsection{Case-I}
These texture five zero quark mass matrices with a minimal phase structure are expressed below
\begin{eqnarray}
{M_{\rm{u}}} = \left( {\begin{array}{*{20}{c}}
\bf{0}&\bf{0}&{f_{\rm u}}e^{-i\gamma_{u}}\\
\bf{0}&{{d_{\rm{u}}}}&\bf{0}\\
{f_{\rm u}}e^{i\gamma_{u}}&\bf{0}&{{c_{\rm{u}}}}
\end{array}} \right)
\end{eqnarray}
or
\begin{eqnarray}
{M_{\rm{u}}} = \left( {\begin{array}{*{20}{c}}
\bf{0}&\bf{0}&{f_{\rm u}}e^{-i\gamma_{u}}\\
\bf{0}&{{d_{\rm{u}}}}&\bf{0}\\
{f_{\rm u}}e^{-i\gamma_{u}}&\bf{0}&{{c_{\rm{u}}}}
\end{array}} \right),\;\nonumber\\
{M_{\rm{d}}} = \left( {\begin{array}{*{20}{c}}
\bf{0}&{{a_{\rm{d}}}}&\bf{0}\\
{{a_{\rm{d}}}}&{{d_{\rm{d}}}}&{{b_d}}\\
\bf{0}&{{b_d}}&{{c_{\rm{d}}}}
\end{array}} \right),
\end{eqnarray}
where  
\begin{eqnarray}
{f_{\rm u}}=\sqrt{m_{\rm u}m_{\rm t}},\;\nonumber\\
d_{\rm u}=m_{\rm c},\;\nonumber\\
c_{\rm u}=m_{\rm t}-m_{\rm u}\;\nonumber\\
{c_{\rm d}} =  - {m_{\rm d}} + {m_{\rm s}} + {m_{\rm b}} - {d_{\rm d}},\;\nonumber\\
a_{\rm d} = \sqrt {{{{m_{\rm{d}}}{m_{\rm{s}}}{m_{\rm{b}}}} \mathord{\left/
 {\vphantom {{{m_{\rm{d}}}{m_{\rm{s}}}{m_{\rm{b}}}} {{c_{\rm{d}}}}}} \right.
 \kern-\nulldelimiterspace} {{c_{\rm{d}}}}}} ,\;\nonumber\\
{b_{\rm{d}}} = \sqrt {{{({c_{\rm{d}}} + {m_{\rm{d}}})({c_{\rm{d}}} - {m_{\rm{s}}})({m_{\rm{b}}} - {c_{\rm{d}}})} \mathord{\left/
 {\vphantom {{({c_{\rm{d}}} + {m_{\rm{d}}})({c_{\rm{d}}} - {m_{\rm{s}}})({m_{\rm{b}}} - {c_{\rm{d}}})} {{c_{\rm{d}}}}}} \right.
 \kern-\nulldelimiterspace} {{c_{\rm{d}}}}}}\nonumber \\ \end{eqnarray}
involving only two free parameters $d_{\rm d}$ and $\gamma_{u}$. Clearly, predictions are expected from these mass matrices. The resulting quark mixing matrix is obtained using $V=O^{T}_{\rm u}P O_{\rm d}$, where $P=diag\lbrace e^{i\gamma_{\rm u}},1,1\rbrace $ such that
\begin{eqnarray}
O_{\rm{u}} = \left( {\begin{array}{*{20}{c}}
{\sqrt {\frac{{{m_{\rm{t}}}}}{{({m_{\rm{t}}} + {m_{\rm{u}}})}}} }&0&{ -\sqrt {\frac{{{m_{\rm{u}}}}}{{({m_{\rm{t}}} + {m_{\rm{u}}})}}} }\\
0&1&0\\
{\sqrt {\frac{{{m_{\rm{u}}}}}{{({m_{\rm{t}}} + {m_{\rm{u}}})}}} }&0&{\sqrt {\frac{{{m_{\rm{t}}}}}{{({m_{\rm{t}}} + {m_{\rm{u}}})}}} }
\end{array}} \right),\nonumber \\
\end{eqnarray}
\begin{widetext}
\begin{eqnarray}
{O_{\rm{d}}} = \left( {\begin{array}{*{20}{c}}
{\sqrt {\frac{{{m_{\rm{s}}}{m_{\rm{b}}}({c_{\rm{d}}} + {m_{\rm{d}}})}}{{{c_{\rm{d}}}({m_{\rm{d}}} + {m_{\rm{s}}})({m_{\rm{b}}} + {m_{\rm{d}}})}}} }&{\sqrt {\frac{{{m_{\rm{d}}}{m_{\rm{b}}}({c_{\rm{d}}} - {m_{\rm{s}}})}}{{{c_{\rm{d}}}({m_{\rm{b}}} - {m_{\rm{s}}})({m_{\rm{d}}} + {m_{\rm{s}}})}}} }&{\sqrt {\frac{{{m_{\rm{d}}}{m_{\rm{s}}}({m_{\rm{b}}} - {c_{\rm{d}}})}}{{{c_{\rm{d}}}({m_{\rm{b}}} - {m_{\rm{s}}})({m_{\rm{d}}} + {m_{\rm{b}}})}}} }\\
{ - \sqrt {\frac{{{m_{\rm{d}}}({c_{\rm{d}}} + {m_{\rm{d}}})}}{{({m_{\rm{d}}} + {m_{\rm{s}}})({m_{\rm{b}}} + {m_{\rm{d}}})}}} }&{\sqrt {\frac{{{m_{\rm{s}}}({c_{\rm{d}}} - {m_{\rm{s}}})}}{{({m_{\rm{b}}} - {m_{\rm{s}}})({m_{\rm{d}}} + {m_{\rm{s}}})}}} }&{\sqrt {\frac{{{m_{\rm{b}}}({m_{\rm{b}}} - {c_{\rm{d}}})}}{{({m_{\rm{b}}} - {m_{\rm{s}}})({m_{\rm{d}}} + {m_{\rm{b}}})}}} }\\
{\sqrt {\frac{{{m_{\rm{d}}}({c_{\rm{d}}} - {m_{\rm{s}}})({m_{\rm{b}}} - {c_{\rm{d}}})}}{{{c_{\rm{d}}}({m_{\rm{d}}} + {m_{\rm{s}}})({m_{\rm{b}}} + {m_{\rm{d}}})}}} }&{ - \sqrt {\frac{{{m_{\rm{s}}}({c_{\rm{d}}} + {m_{\rm{d}}})({m_{\rm{b}}} - {c_{\rm{d}}})}}{{{c_{\rm{d}}}({m_{\rm{b}}} - {m_{\rm{s}}})({m_{\rm{d}}} + {m_{\rm{s}}})}}} }&{\sqrt {\frac{{{m_{\rm{b}}}({c_{\rm{d}}} + {m_{\rm{d}}})({c_{\rm{d}}} - {m_{\rm{s}}})}}{{{c_{\rm{d}}}({m_{\rm{b}}} - {m_{\rm{s}}})({m_{\rm{d}}} + {m_{\rm{b}}})}}} }
\end{array}} \right),
\end{eqnarray}
\end{widetext}
\begin{equation}
V = \left( {\begin{array}{*{20}{c}}
{\sqrt {1 - {{\left| {{V_{us}}} \right|}^2}} }&{\left| {{V_{us}}} \right|}&{  \left| {{V_{ub}}} \right|{e^{ - i{\gamma_{u} }}}}\\
{ - |{V_{{{us}}}}||{V_{{{tb}}}}|}&{|{V_{{{ud}}}}||{V_{{{tb}}}}|}&{\left| {{V_{cb}}} \right|}\\
|{V_{{{td}}}}|{e^{-i{\beta}}}&-{  |{V_{{{cb}}}}||{V_{{{ud}}}}|}&{\sqrt {1 - {{\left| {{V_{cb}}} \right|}^2}} }
\end{array}} \right).
\end{equation}
Using $m_1<<m_2\sim d_{q}<<m_3$, one obtains
\begin{eqnarray}
|{V_{cb}}| =s_{23}=s^{d}_{23}= \sqrt {{{{d_{\rm{d}}} - {m_{\rm{s}}} + {m_{\rm{d}}}} \mathord{\left/
 {\vphantom {{{d_{\rm{d}}} - {m_{\rm{s}}} + {m_{\rm{d}}}} {{m_b}}}} \right.
 \kern-\nulldelimiterspace} {{m_b}}}} ,\;\nonumber\\
V_{td}=s^{d}_{12}s^{d}_{23}-c^{d}_{12}s^{u}_{13}e^{i\gamma_{u}},\;\nonumber\\
\beta=-Arg\lbrace V_{td}\rbrace. \nonumber \\
\end{eqnarray}
It may be noted that the flavor mixing angles are independent of the quark mass  $m_{\rm c}$ and the following predictions emerge within the valid experimental bounds
\begin{eqnarray}
|V_{us}|=s_{12}=s^{d}_{12}\simeq{\sqrt {\frac{{{m_d}}}{{{m_s} + {m_d}}}} },\;\nonumber\\
V_{ub}=s_{13}e^{-i\delta_{13}}\simeq s^{u}_{13}e^{-i\gamma_{u}}={  \sqrt {\frac{{{m_u} }}{{{m_t}}}} {e^{ - i{\gamma _u}}}},\;\nonumber\\
\delta_{13}\simeq \gamma_{u}.\nonumber \\
\end{eqnarray}
With $d_d \simeq m_s$, one observes $V_{cb}\sim \sqrt{m_d/m_b}$ for this structure. The best-fit values for $V$ along with the various CP-angles and $J_{\rm{CP}}$ appear below, e.g.
\begin{eqnarray}
|V| = \left( {\begin{array}{*{20}{c}}
{0.974337}&{0.225065}&0.003569\\
{0.224922}&{0.973508}&{0.041122}\\
0.008781 &{0.040332}&{0.999147}
\end{array}} \right), \nonumber\\
\delta_{13}=71.30^{\circ},~~ 
\phi=-22.05^{\circ},\nonumber\\
J_{CP}=3.046\times 10^{-5},\nonumber\\
\alpha=86.68^{\circ},~~\beta=22.05^{\circ},~~\gamma=71.27^{\circ}\nonumber \\
\end{eqnarray}
which correspond to
\begin{eqnarray}
\gamma_{u}=75.09^{\circ},~~
{m_{\rm{u}}} = 2.11~{\rm{MeV}},~{m_{\rm{t}}} = 172.1~{\rm{GeV}},\nonumber\\
{m_{\rm{d}}} = 3.95~{\rm{MeV}},~ {m_{\rm{s}}} = 74.0~{\rm{MeV}},~{m_{\rm{b}}} = 2.86~{\rm{GeV}}.\nonumber \\
\end{eqnarray}
All of these are in excellent agreement with the current precision measurement data on quark masses and flavor mixing. The corresponding quark mass matrices (in units of GeV) are 
\begin{eqnarray}
\mid{M_{\rm{u}}}\mid = \left( {\begin{array}{*{20}{c}}
{\bf{0}}&{\bf{0}}&0.602603\\
{\bf{0}}&{m_{\rm c}}&{\bf{0}}\\
0.602603&{\bf{0}}&{172.09789}
\end{array}} \right), \nonumber\\
{M_{\rm{d}}} = \left( {\begin{array}{*{20}{c}}
{\bf{0}}&{0.017110}&{\bf{0}}\\
{0.017110}&{0.074767}&{0.114628}\\
{\bf{0}}&{0.114628}&{2.855282}
\end{array}} \right).
\end{eqnarray}
It is observed that the mixing parameters $\mid V_{ub}\mid$ and $sin~2\beta$ are highly sensitive to the quark mass $m_{u}>2.0~MeV$ for viability, as depicted in FIG.~1 and FIG.~2 respectively.
\begin{figure}
\includegraphics[scale=1.0]{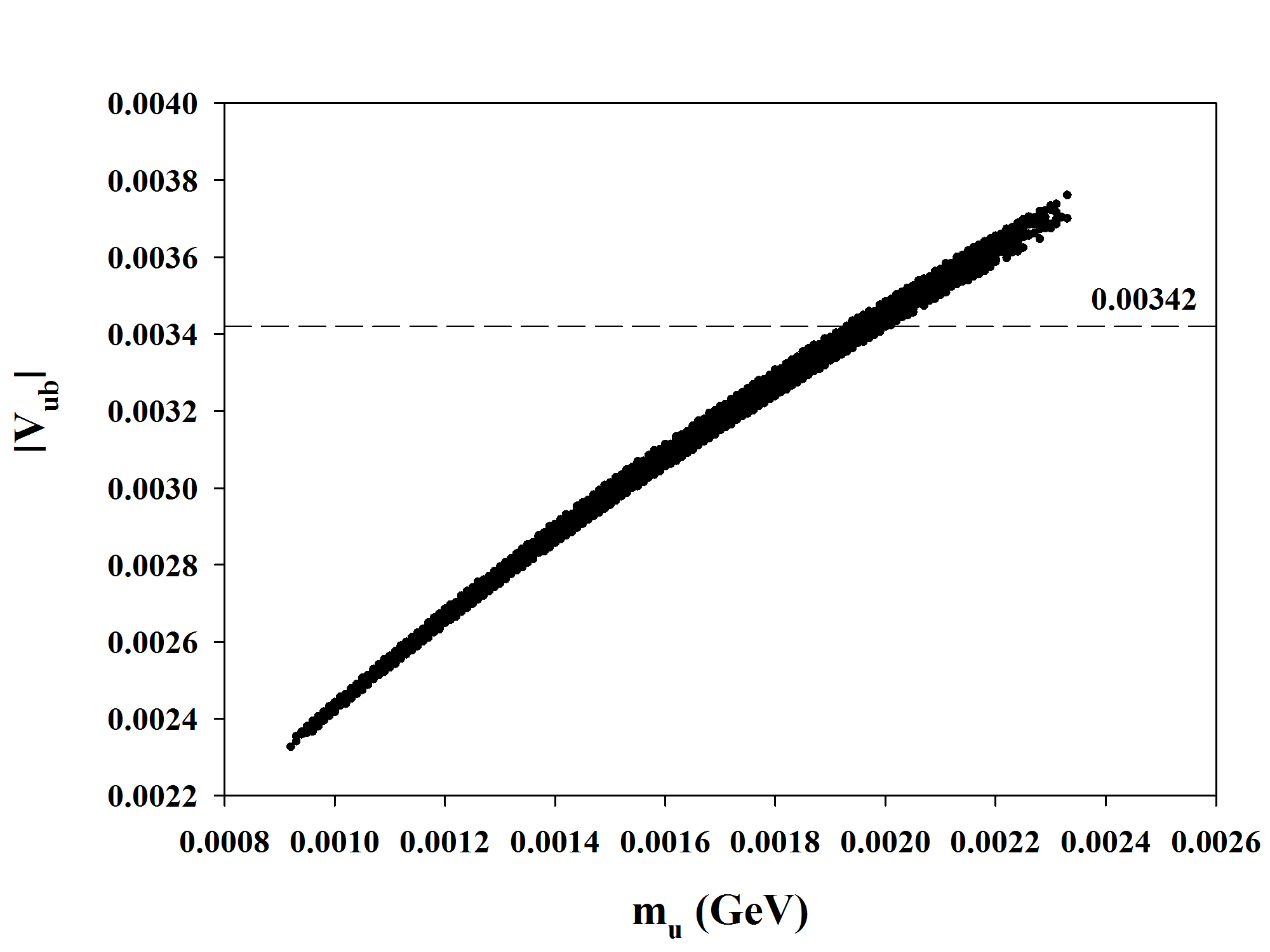}
\caption{$\mid V_{ub}\mid$ vs. $m_{u}$ for Case-I}
\end{figure}
\begin{figure}
\includegraphics[scale=1.0]{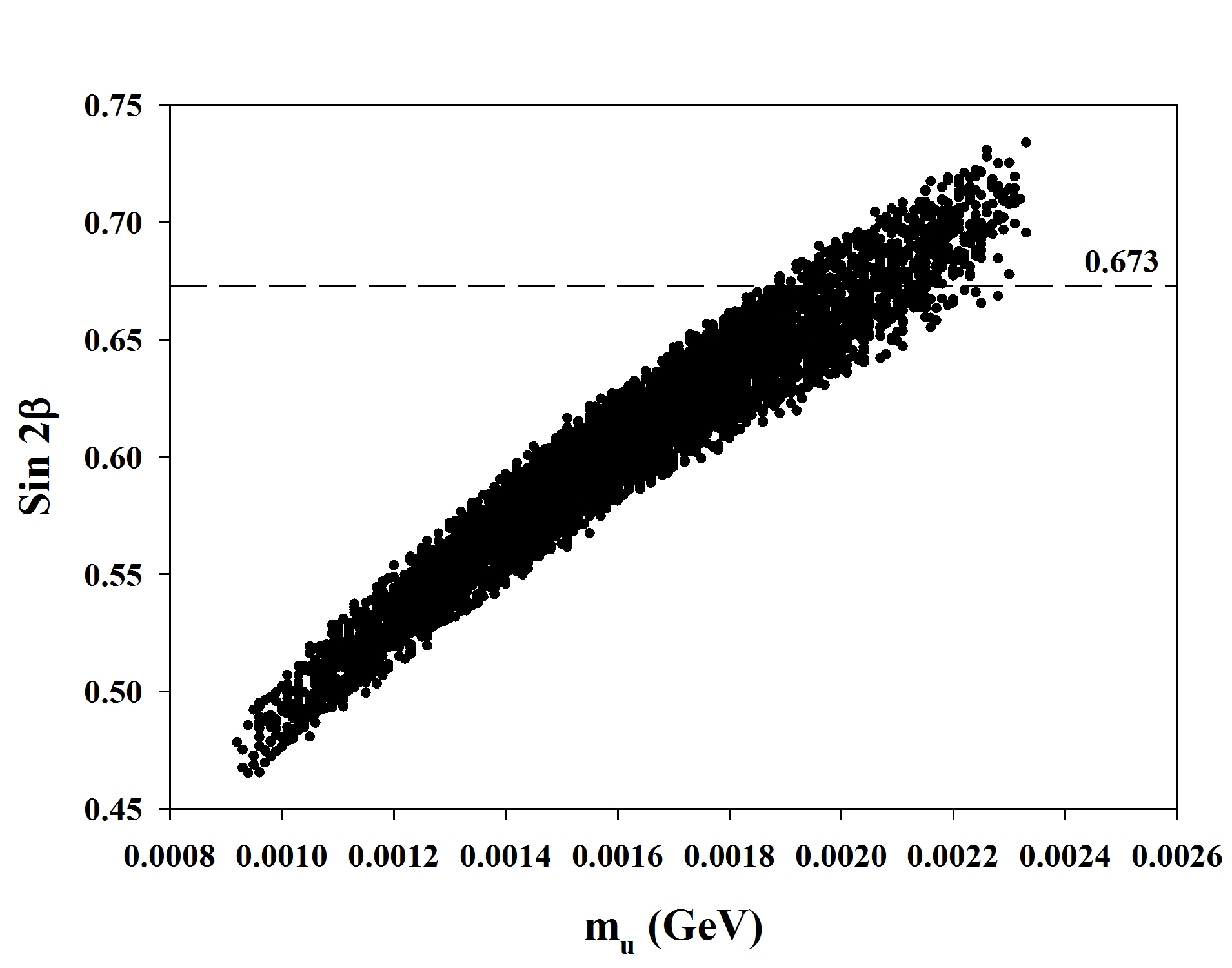}
\caption{$sin~2\beta$ vs. $m_{u}$ for Case-I}
\end{figure}
\begin{figure}
\includegraphics[scale=1.0]{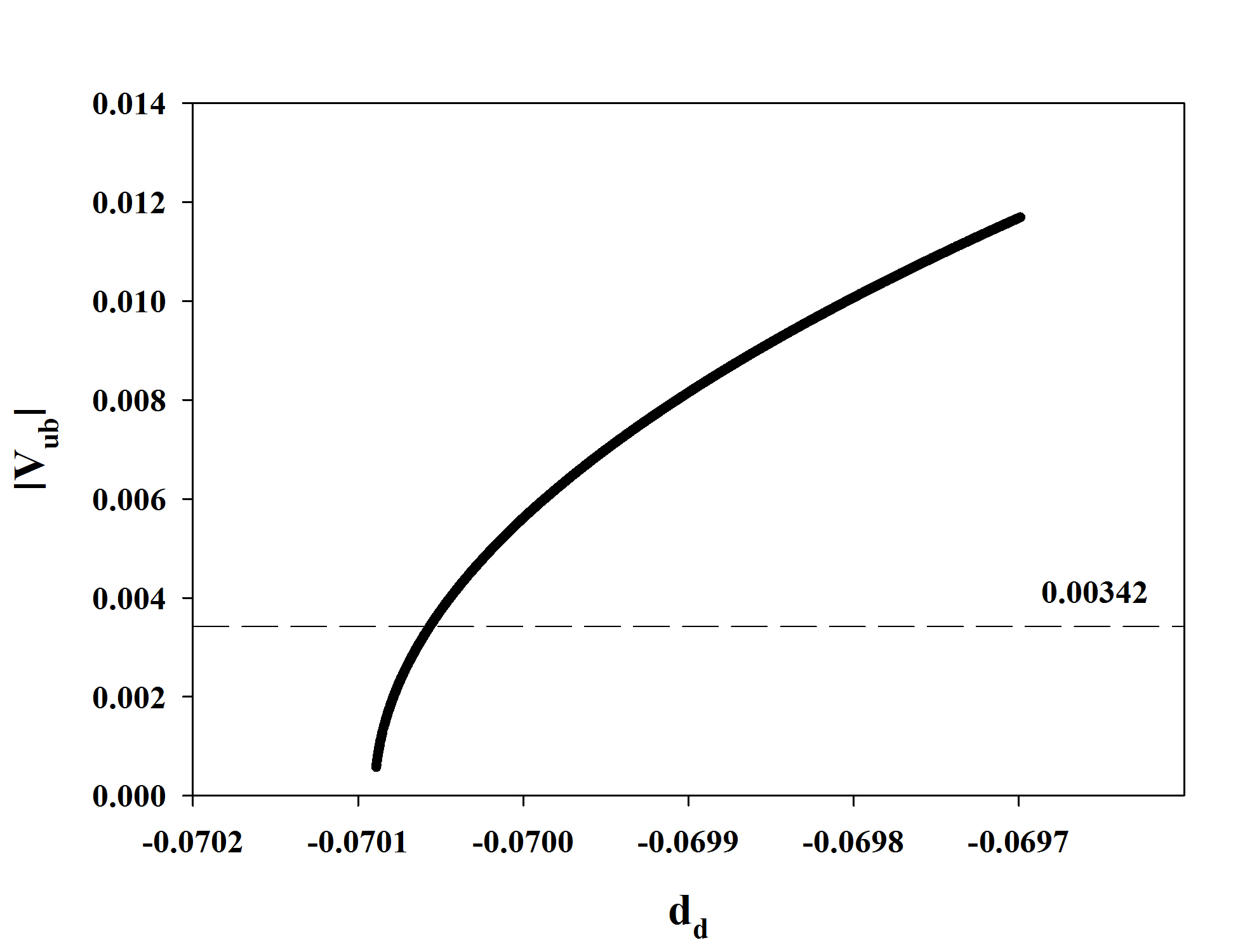}
\caption{$\mid V_{ub}\mid$ vs. $d_{d}$ for Case-II}
\end{figure}
\begin{figure}
\includegraphics[scale=1.0]{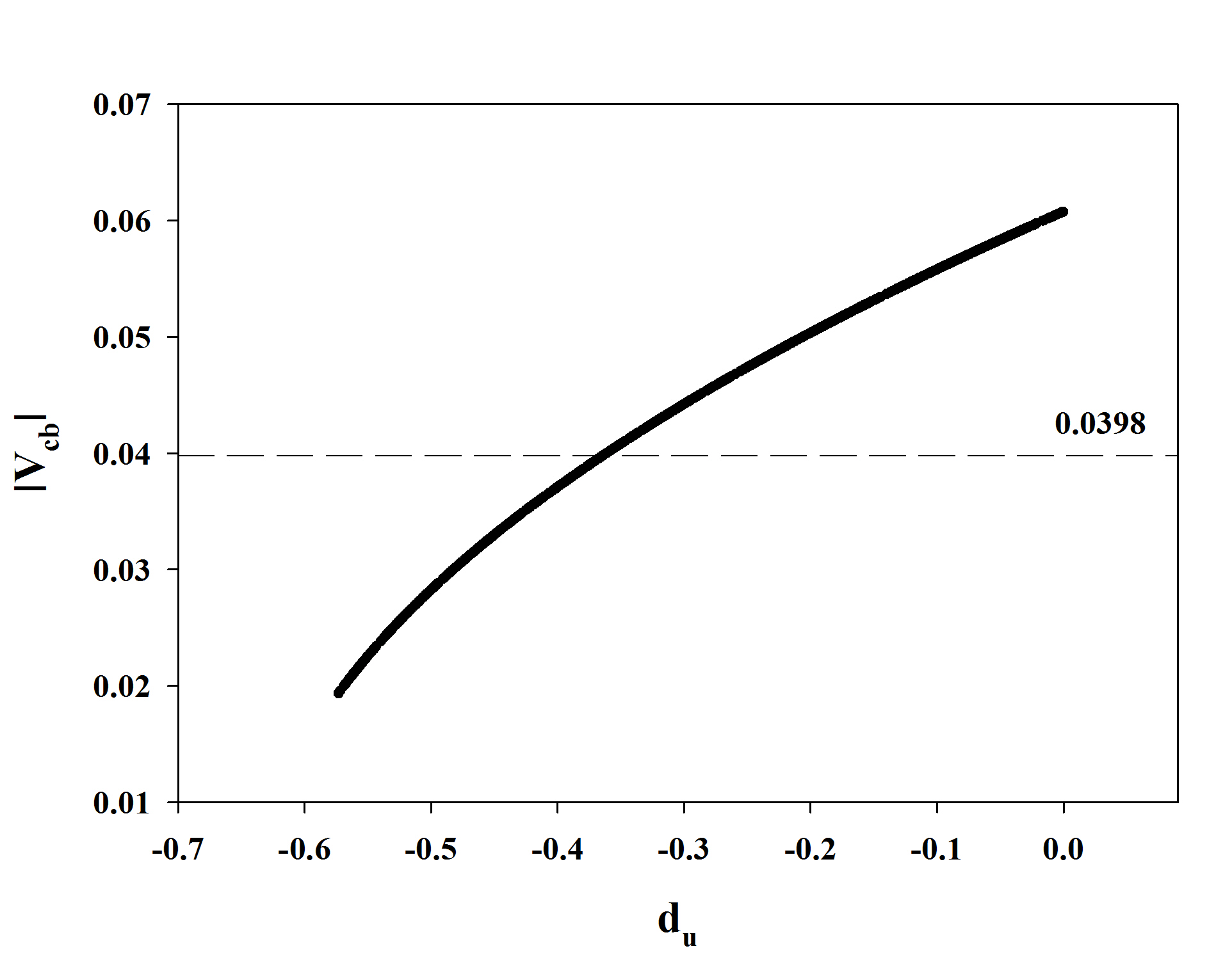}
\caption{$\mid V_{cb}\mid$ vs. $d_{u}$ for Case-II}
\end{figure}
\subsection{Case-II}
These texture four zero quark mass matrices with a minimal phase structure are expressed below
\begin{eqnarray}
{M_{\rm{u}}} = \left( {\begin{array}{*{20}{c}}
{{m_u}}&0&0\\
0&{{d_{\rm{u}}}}&{{b_u}}\\
0&{{b_u}}&{{c_{\rm{u}}}}
\end{array}} \right),\\
{M_{\rm{d}}} = \left( {\begin{array}{*{20}{c}}
0&{{a_{\rm{d}}}}&{{f_d}{e^{ - i{\gamma _d}}}}\\
{{a_{\rm{d}}}}&{{d_{\rm{d}}}}&0\\
{{f_d}{e^{  i{\gamma _d}}}}&0&{{c_{\rm{d}}}}
\end{array}} \right)\;\nonumber
\end{eqnarray}
or
\begin{eqnarray}
{M_{\rm{d}}} = \left( {\begin{array}{*{20}{c}}
0&{{a_{\rm{d}}}}&{{f_d}{e^{- i{\gamma _d}}}}\\
{{a_{\rm{d}}}}&{{d_{\rm{d}}}}&0\\
{{f_d}{e^{ - i{\gamma _d}}}}&0&{{c_{\rm{d}}}}{e^{- 2i{\gamma _d}}}
\end{array}} \right),
\end{eqnarray}
\begin{eqnarray}
{b_{\rm u}} = \sqrt {({m_{\rm{t}}} - {d_{\rm{u}}})({m_{\rm{c}}} + {d_{\rm{u}}})}  ,\;\nonumber\\
{c_{\rm{u}}} =  - {m_{\rm{c}}} + {m_{\rm{t}}} - {d_{\rm{u}}},\;\nonumber\\
{c_{\rm{d}}} = {m_{\rm{d}}} - {m_{\rm{s}}} + {m_{\rm{b}}} - {d_{\rm{d}}},\;\nonumber\\
{a_{\rm{d}}} = \sqrt {\frac{{({m_{\rm{d}}} - {d_{\rm{d}}})({m_{\rm{s}}} + {d_{\rm{d}}})({m_{\rm{b}}} - {d_{\rm{d}}})}}{{({c_{\rm{d}}} - {d_{\rm{d}}})}}},\;\nonumber\\
{f_{\rm{d}}} = \sqrt {\frac{{({c_{\rm{d}}} - {m_{\rm{d}}})({m_{\rm{s}}} + {c_{\rm{d}}})({m_{\rm{b}}} - {c_{\rm{d}}})}}{{({c_{\rm{d}}} - {d_{\rm{d}}})}}}\;\nonumber\\
\end{eqnarray}
involving three free parameters $d_{\rm u}$, $d_{\rm d}$ and $\gamma_{\rm d}$. Again, a few predictions are expected from these matrices. This leads to $V=O^{T}_{\rm u}PO_{\rm d}$, where $P=diag\lbrace 1,1,e^{i\gamma_{\rm d}}\rbrace $ such that
\begin{eqnarray}
{O_{\rm{u}}} = \left( {\begin{array}{*{20}{c}}
1&0&0\\
0&{ - \sqrt {\frac{{\left( {{m_{\rm{t}}} - {d_{\rm{u}}}} \right)}}{{({m_{\rm{t}}} + {m_{\rm{c}}})}}} }&{\sqrt {\frac{{\left( {{d_{\rm{u}}} + {m_{\rm{c}}}} \right)}}{{({m_{\rm{t}}} + {m_{\rm{c}}})}}} }\\
0&{\sqrt {\frac{{\left( {{d_{\rm{u}}} + {m_{\rm{c}}}} \right)}}{{({m_{\rm{t}}} + {m_{\rm{c}}})}}} }&{\sqrt {\frac{{\left( {{m_{\rm{t}}} - {d_{\rm{u}}}} \right)}}{{({m_{\rm{t}}} + {m_{\rm{c}}})}}} }
\end{array}} \right),
\end{eqnarray}
\begin{widetext}
\begin{eqnarray}
{O_{\rm{d}}} = \left( {\begin{array}{*{20}{c}}
{\sqrt {\frac{{({m_{\rm{d}}} - {d_{\rm{d}}})({c_{\rm{d}}} - {m_{\rm{d}}})}}{{({m_{\rm{d}}} + {m_{\rm{s}}})({m_{\rm{b}}} - {m_{\rm{d}}})}}} }&{\sqrt {\frac{{({m_{\rm{s}}} + {d_{\rm{d}}})({c_{\rm{d}}} + {m_{\rm{s}}})}}{{({m_{\rm{d}}} + {m_{\rm{s}}})({m_{\rm{b}}} + {m_{\rm{s}}})}}} }&{\sqrt {\frac{{({m_{\rm{b}}} - {d_{\rm{d}}})({m_{\rm{b}}} - {c_{\rm{d}}})}}{{({m_{\rm{b}}} + {m_{\rm{s}}})({m_{\rm{b}}} - {m_{\rm{d}}})}}} }\\
{\sqrt {\frac{{({m_{\rm{s}}} + {d_{\rm{d}}})({m_{\rm{b}}} - {d_{\rm{d}}})({c_{\rm{d}}} - {m_{\rm{d}}})}}{{({c_{\rm{d}}} - {d_{\rm{d}}})({m_{\rm{d}}} + {m_{\rm{s}}})({m_{\rm{b}}} - {m_{\rm{d}}})}}} }&{ - \sqrt {\frac{{({m_{\rm{d}}} - {d_{\rm{d}}})({m_{\rm{b}}} - {d_{\rm{d}}})({c_{\rm{d}}} + {m_{\rm{s}}})}}{{({c_{\rm{d}}} - {d_{\rm{d}}})({m_{\rm{d}}} + {m_{\rm{s}}})({m_{\rm{b}}} + {m_{\rm{s}}})}}} }&{\sqrt {\frac{{({m_{\rm{d}}} - {d_{\rm{d}}})({m_{\rm{s}}} + {d_{\rm{d}}})({m_{\rm{b}}} - {c_{\rm{d}}})}}{{({c_{\rm{d}}} - {d_{\rm{d}}})({m_{\rm{b}}} + {m_{\rm{s}}})({m_{\rm{b}}} - {m_{\rm{d}}})}}} }\\
{ - \sqrt {\frac{{({m_{\rm{d}}} - {d_{\rm{d}}})({c_{\rm{d}}} + {m_{\rm{s}}})({m_{\rm{b}}} - {c_{\rm{d}}})}}{{({c_{\rm{d}}} - {d_{\rm{d}}})({m_{\rm{d}}} + {m_{\rm{s}}})({m_{\rm{b}}} - {m_{\rm{d}}})}}} }&{ - \sqrt {\frac{{({m_{\rm{s}}} + {d_{\rm{d}}})({c_{\rm{d}}} - {m_{\rm{d}}})({m_{\rm{b}}} - {c_{\rm{d}}})}}{{({c_{\rm{d}}} - {d_{\rm{d}}})({m_{\rm{d}}} + {m_{\rm{s}}})({m_{\rm{b}}} + {m_{\rm{s}}})}}} }&{\sqrt {\frac{{({m_{\rm{b}}} - {d_{\rm{d}}})({c_{\rm{d}}} - {m_{\rm{d}}})({c_{\rm{d}}} + {m_{\rm{s}}})}}{{({c_{\rm{d}}} - {d_{\rm{d}}})({m_{\rm{b}}} + {m_{\rm{s}}})({m_{\rm{b}}} - {m_{\rm{d}}})}}} }
\end{array}} \right),
\end{eqnarray}
\end{widetext}
\begin{equation}
V = \left( {\begin{array}{*{20}{c}}
{\sqrt {1 - {{\left| {{V_{us}}} \right|}^2}} }&{\left| {{V_{us}}} \right|}&{  \left| {{V_{ub}}} \right|{e^{ - i{\gamma_{d} }}}}\\
{ - |{V_{{{us}}}}||{V_{{{tb}}}}|}&{|{V_{{{ud}}}}||{V_{{{tb}}}}|}&{\left| {{V_{cb}}} \right|}\\
|{V_{{{td}}}}|{e^{-i{\beta}}}&-{  |{V_{{{cb}}}}||{V_{{{ud}}}}|}&{\sqrt {1 - {{\left| {{V_{cb}}} \right|}^2}} }
\end{array}} \right).
\end{equation}

Using $m_1<<m_2\sim d_{q}<<m_3$, one obtains
\begin{eqnarray}\label{case3 rule out}
|V_{us}|={s_{12}} = s_{12}^d = \sqrt {\frac{{{m_s} + {d_d}}}{{{m_s} + {m_d}}}} ,\;\nonumber\\
V_{ub}=s^{d}_{13}e^{-i\gamma_{d}}={\sqrt {\frac{{{d_d} - {m_d} + {m_s}}}{{{m_b}}}} }e^{-i\gamma_{d}},\;\nonumber\\
|V_{cb}|=s_{23}=s^{u}_{23}={\sqrt {\frac{{{d_u} + {m_c}}}{{{m_t} + {m_c}}}} },\;\nonumber\\
V_{td}=s^{d}_{12}s^{u}_{23}-c^{d}_{12}s^{d}_{13}e^{i\gamma_{d}},\;\nonumber\\
\delta_{13}=\gamma_{d},\;\nonumber\\
\beta=-Arg\lbrace V_{td}\rbrace.
\end{eqnarray}
The best-fit values for $V$ along with the various CP-angles and $J_{\rm{CP}}$ appear below, e.g.
\begin{eqnarray}
|V| = \left( {\begin{array}{*{20}{c}}
{0.974337}&{0.225062}&0.003569\\
{0.224906}&{0.973509}&{0.041173}\\
0.009092 &{0.040315}&{0.999145}
\end{array}} \right), \nonumber\\
\delta_{13}=76.28^{\circ},~~
\phi=-21.82^{\circ},\nonumber\\
J_{CP}=3.127\times 10^{-5},\nonumber\\
\alpha=81.93^{\circ},~~\beta=21.82^{\circ},~~\gamma=76.25^{\circ}
\end{eqnarray}
which correspond to
\begin{eqnarray}
\gamma_{d}=76.25^{\circ},\nonumber\\
{m_{\rm{c}}} = 0.638~{\rm{GeV}},~{m_{\rm{t}}} = 172.1~{\rm{GeV}},\nonumber\\
{m_{\rm{d}}} = 3.91~{\rm{MeV}},~ {m_{\rm{s}}} = 74.0~{\rm{MeV}},~{m_{\rm{b}}} = 2.86~{\rm{GeV}}.\nonumber \\
\end{eqnarray}
All of these are in excellent agreement with the current precision measurement data on quark masses and flavor mixing. The corresponding quark mass matrices (in units of GeV) are
\begin{eqnarray}
{M_{\rm{u}}} = \left( {\begin{array}{*{20}{c}}
m_{\rm u}&{\bf{0}}&{\bf{0}}\\
{\bf{0}}&{-0.34509}&7.10710\\
{\bf{0}}&7.10710&{171.80709}
\end{array}} \right),\nonumber \\
\mid{M_{\rm{d}}}\mid = \left( {\begin{array}{*{20}{c}}
{\bf{0}}&{0.01708}&{0.01020}\\
{0.01708}&{-0.070053}&{\bf{0}}\\
{0.1020}&{\bf{0}}&{2.85996}
\end{array}} \right).
\end{eqnarray}

Interestingly, one observes that $d_{d}=m_{d}-m_{s}+\epsilon \simeq m_{d}-m_{s}$, with $\epsilon<<m_{d}$ such that $s_{13}=\sqrt{\epsilon/m_{b}}$. It may be noted that the flavor mixing angles are independent of the quark mass  $m_{\rm u}$ and the following predictions emerge within the current experimental bounds
\begin{eqnarray}
|V_{us}|={s_{12}} = s_{12}^d = \sqrt {\frac{{{m_d}}}{{{m_s} + {m_d}}}},\nonumber\\
\delta_{13}=\gamma_{d},~
\beta=-Arg\lbrace V_{td}\rbrace.
\end{eqnarray} 
FIGs. 3 and 4 depict the dependence of $\mid V_{ub}\mid$ on $d_{d}$ and of $\mid V_{cb}\mid$ on $d_{u}$ respectively. Additionally, the predictability is enhanced for $d_u=-m_c/2$ in agreement with current precision data at the level of 1$\sigma$ yielding an additional relation ${V_{cb}} = \sqrt {{{{m_c}} \mathord{\left/
 {\vphantom {{{m_c}} {2\left( {{m_t} + {m_c}} \right)}}} \right.
 \kern-\nulldelimiterspace} {2\left( {{m_t} + {m_c}} \right)}}} $. From Eq.(\ref{case3 rule out}), it is observed that the texture five zero possibility with textures interchanged in Case-II for $M_{u}$ and $M_{d}$ are not compatible for $V_{ub}$ and $V_{us}$ simultaneously, since $\sqrt{m_{u}/m_{c}}<<\mid V_{us}\mid$ for $d_{u}\sim m_{c}$ on account of stronger mass hierarchy in the 'up' quark sector. Therefore, the Case-III is ruled out by the current data.
\subsection{Case-IV}
These texture five zero quark mass matrices with a minimal phase structure are expressed below
\begin{eqnarray}
{M_{\rm{u}}} = \left( {\begin{array}{*{20}{c}}
0&0&{{f_u}}\\
0&{{d_{\rm{u}}}}&{{b_{\rm{u}}}}\\
{{f_u}}&{{b_{\rm{u}}}}&{{c_{\rm{u}}}}
\end{array}} \right),\nonumber \\
\end{eqnarray}
and
\begin{eqnarray}
{M_{\rm{d}}} = \left( {\begin{array}{*{20}{c}}
0&{{a_{\rm{d}}}{e^{i{\alpha_{d}}}}}&0\\
{{a_{\rm{d}}}{e^{-i{\alpha_{d}}}}}&{{d_{\rm{d}}}}&0\\
0&0&{{c_{\rm{d}}}}
\end{array}} \right),\;\nonumber
\end{eqnarray}
or
\begin{eqnarray}
{M_{\rm{d}}} = \left( {\begin{array}{*{20}{c}}
0&{{a_{\rm{d}}}{e^{-i{\alpha_{d}}}}}&0\\
{{a_{\rm{d}}}{e^{-i{\alpha_{d}}}}}&{{d_{\rm{d}}}}&0\\
0&0&{{c_{\rm{d}}}}
\end{array}} \right)
\end{eqnarray}
where 
\begin{eqnarray}
{a_d} = \sqrt {{m_d}{m_s}},~~
{d_d} = {m_d} - {m_s},~~
{c_d} = {m_b},\;\nonumber\\
{{f_u}}= \sqrt {\frac{{{m_{\rm u}}{m_{\rm c}}{m_{\rm t}}}}{{{d_{\rm u}}}}},\;\nonumber\\
b_{\rm u} = \sqrt {\frac{{({d_{\rm u}} + {m_{\rm u}})({d_{\rm u}} - {m_{\rm c}})({m_{\rm t}} - {d_{\rm u}})}}{{{d_{\rm u}}}}},\;\nonumber\\
{c_{\rm u}} =  - {m_u} + {m_c} + {m_t}-d_u\;\nonumber\\
\end{eqnarray}
involving two free parameters $d_{\rm u}$ and $\alpha_{d}$. Also the flavor mixing is independent of the quark mass $m_{\rm b}$. As a result, there are only seven parameters involved in these and predictions are again expected. The resulting quark mixing matrix $V=O^{T}_{\rm u}PO_{\rm d}$, where $P=diag\lbrace e^{i\alpha_{d}},1,1\rbrace $ and 
\begin{widetext}
\begin{eqnarray}
{O_{\rm d}} = \left( {\begin{array}{*{20}{c}}
{\sqrt {\frac{{{m_s}}}{{({m_d} + {m_s})}}} }&{\sqrt {\frac{{{m_d}}}{{({m_d} + {m_s})}}} }&0\\
{\sqrt {\frac{{{m_d}}}{{({m_d} + {m_s})}}} }&{ - \sqrt {\frac{{{m_s}}}{{({m_d} + {m_s})}}} }&0\\
0&0&1
\end{array}} \right),\\
{O_{\rm{u}}} = \left( {\begin{array}{*{20}{c}}
{\sqrt {\frac{{{m_{\rm{c}}}{m_{\rm{t}}}({d_{\rm{u}}} + {m_{\rm{u}}})}}{{{d_{\rm{u}}}({m_{\rm{u}}} + {m_{\rm{c}}})({m_{\rm{t}}} + {m_{\rm{u}}})}}} }&{\sqrt {\frac{{{m_{\rm{u}}}{m_{\rm{t}}}({d_{\rm{u}}} - {m_{\rm{c}}})}}{{{d_{\rm{u}}}({m_{\rm{t}}} - {m_{\rm{c}}})({m_{\rm{u}}} + {m_{\rm{c}}})}}} }&{\sqrt {\frac{{{m_{\rm{u}}}{m_{\rm{c}}}({m_{\rm{t}}} - {d_{\rm{u}}})}}{{{d_{\rm{u}}}({m_{\rm{t}}} - {m_{\rm{c}}})({m_{\rm{u}}} + {m_{\rm{t}}})}}} }\\
{\sqrt {\frac{{{m_{\rm{u}}}({d_{\rm{u}}} - {m_{\rm{c}}})({m_{\rm{t}}} - {d_{\rm{u}}})}}{{{d_{\rm{u}}}({m_{\rm{u}}} + {m_{\rm{c}}})({m_{\rm{t}}} + {m_{\rm{u}}})}}} }&{ - \sqrt {\frac{{{m_{\rm{c}}}({d_{\rm{u}}} + {m_{\rm{u}}})({m_{\rm{t}}} - {d_{\rm{u}}})}}{{{d_{\rm{u}}}({m_{\rm{t}}} - {m_{\rm{c}}})({m_{\rm{u}}} + {m_{\rm{c}}})}}} }&{\sqrt {\frac{{{m_{\rm{t}}}({d_{\rm{u}}} + {m_{\rm{u}}})({d_{\rm{u}}} - {m_{\rm{c}}})}}{{{d_{\rm{u}}}({m_{\rm{t}}} - {m_{\rm{c}}})({m_{\rm{u}}} + {m_{\rm{t}}})}}} }\\
{ - \sqrt {\frac{{{m_{\rm{u}}}({d_{\rm{u}}} + {m_{\rm{u}}})}}{{({m_{\rm{u}}} + {m_{\rm{c}}})({m_{\rm{t}}} + {m_{\rm{u}}})}}} }&{\sqrt {\frac{{{m_{\rm{c}}}({d_{\rm{u}}} - {m_{\rm{c}}})}}{{({m_{\rm{t}}} - {m_{\rm{c}}})({m_{\rm{u}}} + {m_{\rm{c}}})}}} }&{\sqrt {\frac{{{m_{\rm{t}}}({m_{\rm{t}}} - {d_{\rm{u}}})}}{{({m_{\rm{t}}} - {m_{\rm{c}}})({m_{\rm{u}}} + {m_{\rm{t}}})}}} }
\end{array}} \right).
\end{eqnarray}
\end{widetext}
Using $m_1<<m_2\sim d_{q}<<m_3$, one obtains
\begin{eqnarray}\label{below}
V_{ud}={\sqrt {\frac{{{m_s}}}{{{m_d} + {m_s}}}}  + \sqrt {\frac{{{m_u}}}{{{m_c}}} - \frac{{{m_u}}}{{{d_u}}}} \sqrt {\frac{{{m_d}}}{{{m_s} + {m_d}}}} {e^{ - i{\alpha_d}}}}\;\nonumber\\
V_{us}={\sqrt {\frac{{{m_d}}}{{{m_s} + {m_d}}}}  - \sqrt {\frac{{{m_s}}}{{{m_d} + {m_s}}}} \sqrt {\frac{{{m_u}}}{{{m_c}}} - \frac{{{m_u}}}{{{d_u}}}} {e^{ - i{\alpha _d}}}}\;\nonumber\\
V_{ub}={ \sqrt {\frac{{{m_u}}}{{{m_t}}}} \sqrt {\frac{{{d_u}}}{{{m_c}}}}{e^{ - i\left( {{\alpha _d} - \pi } \right)}}}\;\nonumber\\
V_{cd}=\sqrt {\frac{{{m_s}}}{{{m_d} + {m_s}}}} \sqrt {\frac{{{m_u}}}{{{m_c}}} - \frac{{{m_u}}}{{{d_u}}}} {e^{i{\alpha _d}}}\;\nonumber\\
 - \sqrt {\frac{{{m_d}}}{{{m_s} + {m_d}}}} \sqrt {\frac{{{m_t} - {d_u}}}{{{m_t} - {m_c}}}}\;\nonumber\\
 V_{cs}={\sqrt {\frac{{{m_s}}}{{{m_d} + {m_s}}}} \sqrt {\frac{{{m_t} - {d_u}}}{{{m_t} - {m_c}}}} }\;\nonumber\\
 V_{cb}={\sqrt {\frac{{{d_u} - {m_c}}}{{{m_t} - {m_c}}}} }\;\nonumber\\
 V_{td}=\sqrt {\frac{{{m_u}}}{{{m_t}}}} \sqrt {\frac{{{m_c}}}{{{d_u}}}} \sqrt {\frac{{{m_s}}}{{{m_d} + {m_s}}}} {e^{i{\alpha _d}}}\;\nonumber\\
 + \sqrt {\frac{{{m_d}}}{{{m_s} + {m_d}}}} \sqrt {\frac{{{d_u} - {m_c}}}{{{m_t} - {m_c}}}}\;\nonumber\\
 V_{ts}={ - \sqrt {\frac{{{d_u} - {m_c}}}{{{m_t} - {m_c}}}} \sqrt {\frac{{{m_s}}}{{{m_d} + {m_s}}}} }\;\nonumber\\
 V_{tb}={\sqrt {\frac{{{m_t} - {d_u}}}{{{m_t} - {m_c}}}} }.\;\nonumber\\
\end{eqnarray}
\begin{figure}
\includegraphics[scale=1.0]{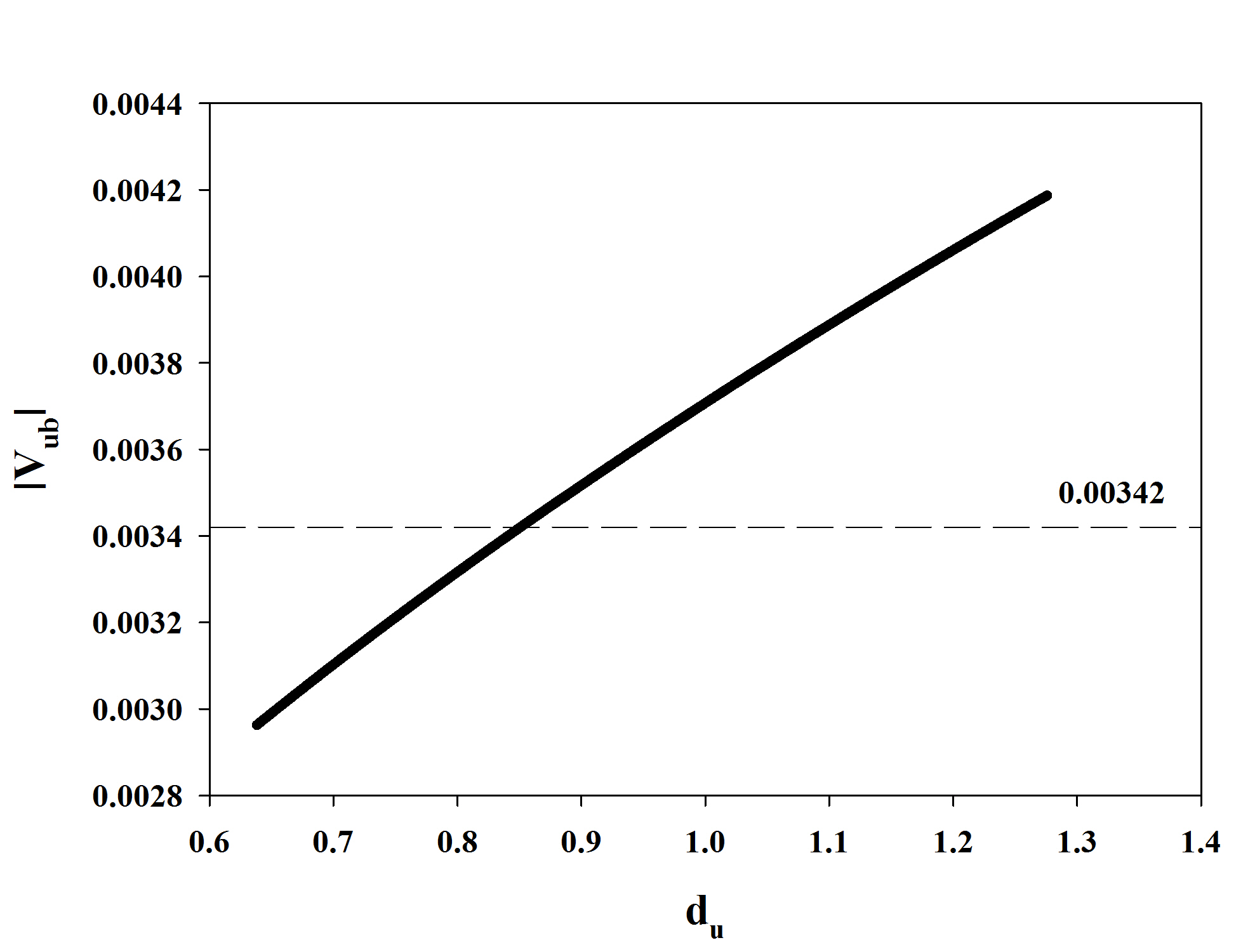}
\caption{$\mid V_{ub}\mid$ vs. $d_{u}$ for Case-IV}
\includegraphics[scale=1.0]{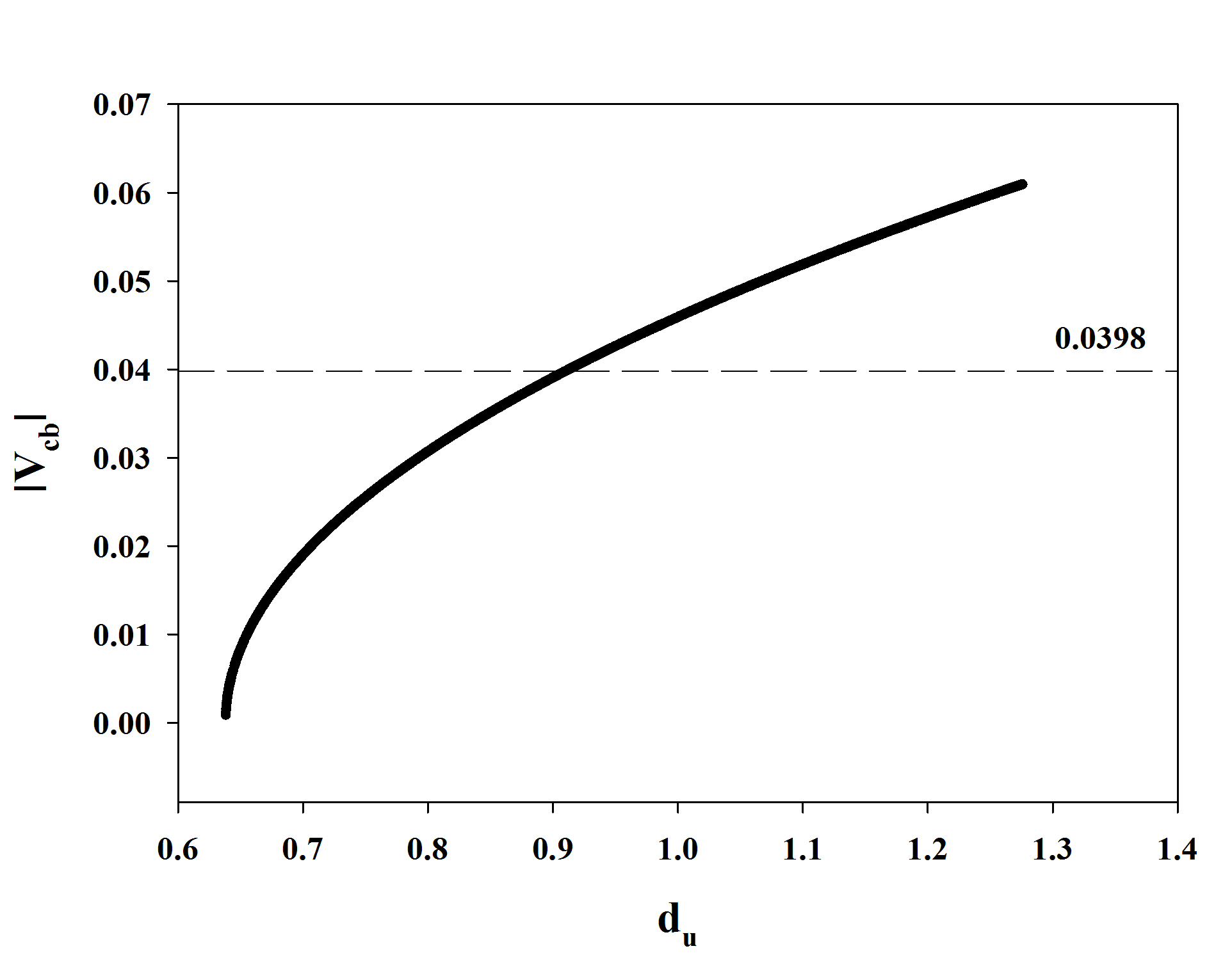}
\caption{$\mid V_{cb}\mid$ vs. $d_{u}$ for Case-IV}
\includegraphics[scale=1.0]{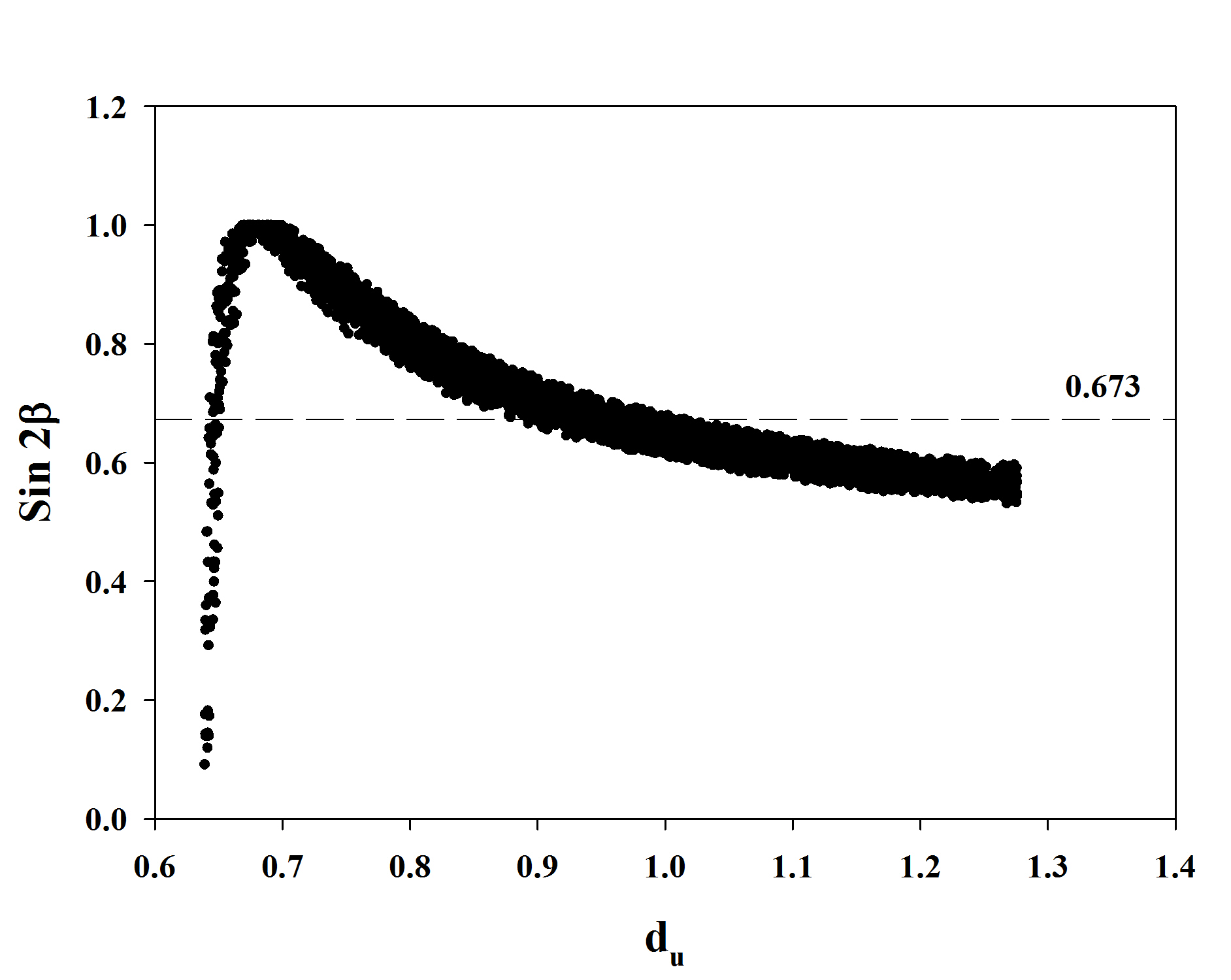}
\caption{$sin~2\beta$ vs. $d_{u}$ for Case-IV}
\end{figure} 
The above relations indicate that the parameter $d_{u}$ contributes non-trivially to all the three flavor mixing angles and that this texture five zero quark mass structure, although appealing at the first look, does not offer any easy predictions for the quark masses and mixing angles. Note that $\beta\neq-Arg\lbrace V_{td}\rbrace$ and one has to rely on the most general definition of $\beta=\arg \left( { - {{V_{cd}^{}V_{cb}^*} \mathord{\left/
 {\vphantom {{V_{cd}^{}V_{cb}^*} {V_{td}^{}V_{tb}^*}}} \right.
 \kern-\nulldelimiterspace} {V_{td}^{}V_{tb}^*}}} \right)$. Also, the approximation ${V_{us}} = \sqrt {{{{m_d}} \mathord{\left/
 {\vphantom {{{m_d}} {({m_s} + {m_d})}}} \right.
 \kern-\nulldelimiterspace} {({m_s} + {m_d})}}}$ deviates from the experimental precision data by approximately 10$\sigma$ and cannot be regarded as a reasonable prediction for such texture. The $d_{u}$ dependence of the other three quark flavor mixing parameters is depicted in FIGs. 5, 6 and 7.

The best-fit values for $V$ along with the various CP-angles and $J_{\rm{CP}}$ appear below, e.g.
\begin{eqnarray}
|V| = \left( {\begin{array}{*{20}{c}}
{0.974338}&{0.225062}&0.003573\\
{0.224905}&{0.973510}&{0.041169}\\
0.009131 &{0.040302}&{0.999145}
\end{array}} \right), \nonumber\\
\delta_{13}=76.93^{\circ},~~ \phi=-15.08^{\circ},~~
J_{CP}=3.14\times 10^{-5},\nonumber\\
\alpha=81.28^{\circ},~~\beta=21.82^{\circ},~~\gamma=76.90^{\circ}.\;\nonumber\\
\end{eqnarray}
These correspond to
\begin{eqnarray}
\alpha_{d}=263.29^{\circ},\;\nonumber\\
{m_{\rm{u}}} = 1.51~{\rm{MeV}},~{m_{\rm{c}}} = 0.638~{\rm{GeV}},~{m_{\rm{t}}} = 172.1~{\rm{GeV}},\nonumber\\
{m_{\rm{d}}} = 2.99~{\rm{MeV}},~ {m_{\rm{s}}} = 58.5~{\rm{MeV}}\;\nonumber\\
\end{eqnarray}
which are also in excellent agreement with the current precision measurement data on quark masses and flavor mixing. The corresponding quark mass matrices (in units of GeV) are
\begin{eqnarray}
{M_{\rm{u}}} = \left( {\begin{array}{*{20}{c}}
{\bf{0}}&{\bf{0}}&0.422388\\
{\bf{0}}&{0.929297}&7.067006\\
0.422388&7.067006&{171.80719}
\end{array}} \right),\;\nonumber\\ \mid{M_{\rm{d}}}\mid = \left( {\begin{array}{*{20}{c}}
{\bf{0}}&{0.013225}&{\bf{0}}\\
{0.013225}&-{0.055510}&{\bf{0}}\\
{\bf{0}}&{\bf{0}}&{m_{\rm b}}
\end{array}} \right). \;\nonumber\\
\end{eqnarray}
However, predictability is greatly increased for $d_{u}=3m_{c}/2$, such that
\begin{eqnarray}
{V_{us}} = \sqrt {\frac{{{m_d}}}{{{m_s} + {m_d}}}}  - \sqrt {\frac{{{m_u}}}{{{m_c}}}} {e^{ - i{\alpha _d}}}\;\nonumber\\
{V_{ub}} = \sqrt {\frac{{3{m_u}}}{{2{m_t}}}} {e^{ - i\left( {{\alpha _d} - \pi } \right)}}\;\nonumber\\
{V_{cb}} = \sqrt {\frac{{{m_c}}}{{2\left( {{m_t} - {m_c}} \right)}}}. 
\end{eqnarray}
The best-fit values for $V$ along with the various CP-angles and $J_{\rm{CP}}$ for $d_{u}=3m_{c}/2$ appear below, e.g.
\begin{eqnarray}
|V| = \left( {\begin{array}{*{20}{c}}
{0.974333}&{0.225084}&0.003504\\
{0.224941}&{0.973537}&{0.040351}\\
0.008750 &{0.039546}&{0.999179}
\end{array}} \right), \nonumber\\
\delta_{13}=73.54^{\circ},~~ \phi=-14.87^{\circ},~~
J_{CP}=2.97\times 10^{-5},\nonumber\\
\alpha=84.51^{\circ},~~\beta=21.99^{\circ},~~\gamma=73.50^{\circ}.\;\nonumber\\\end{eqnarray}
These correspond to
\begin{eqnarray}
\alpha_{d}=260.26^{\circ},\nonumber\\
{m_{\rm{u}}} = 1.41~{\rm{MeV}},~{m_{\rm{c}}} = 0.56~{\rm{GeV}},~{m_{\rm{t}}} = 172.1~{\rm{GeV}},\nonumber\\
{m_{\rm{d}}} = 3.30~{\rm{MeV}},~ {m_{\rm{s}}} = 65.7~{\rm{MeV}}.\;\nonumber\\
\end{eqnarray}
The corresponding quark mass matrices (in units of GeV) are
\begin{eqnarray}
{M_{\rm{u}}} = \left( {\begin{array}{*{20}{c}}
{\bf{0}}&{\bf{0}}&0.402211\\
{\bf{0}}&{0.840000}&6.930061\\
0.402211&6.930061&{171.81859}
\end{array}} \right),\;\nonumber\\ \mid{M_{\rm{d}}}\mid= \left( {\begin{array}{*{20}{c}}
{\bf{0}}&{0.014724}&{\bf{0}}\\
{0.014724}&-{0.062400}&{\bf{0}}\\
{\bf{0}}&{\bf{0}}&{m_{\rm b}}
\end{array}} \right). \nonumber \\
\end{eqnarray}
From Eq.(\ref{below}), it is observed that the texture five zero possibility with textures interchanged in Case-IV for $M_{u}$ and $M_{d}$ is not viable for $V_{ub}$ since $\sqrt{m_{d}/m_{b}}>>\mid V_{ub}\mid$ for $d_{d}\cong m_{s}$. Therefore, the Case-V and Case-VI are ruled out by the current precision data.

To this end, we should also like to comment on the texture five zero structures discussed in \cite{Giraldo:2015cpp, Giraldo:2011ya} characterized by the following texture structure
\begin{equation}
{M_u} = \left( {\begin{array}{*{20}{c}}
0&0& \times \\
0& \times & \times \\
 \times & \times & \times 
\end{array}} \right),{\rm{ }}{M_d} = \left( {\begin{array}{*{20}{c}}
0& \times &0\\
 \times &0& \times \\
0& \times & \times 
\end{array}} \right).
\end{equation}
Our analysis agrees with \cite{Giraldo:2015cpp} and the above structure is also observed to be viable with current precision data. However, the presence of non-zero diagonal elements at (23) and (32) positions in $M_d$ as compared to the Case-IV discussed above signifies an additional LO contribution of rotation $s^{d}_{23}$ to the mixing angle $s_{23}$ as compared to Case-IV, thereby complicating the prediction for $V_{cb}$. Hence, we exclude this structure in our study.
\\
\subsection{Case-VII}
These texture four zero quark mass matrices with a minimal phase structure are expressed below
\begin{eqnarray}
{M_u} = \left( {\begin{array}{*{20}{c}}
e_{u}&{{a_u}{e^{ - i{\alpha _u}}}}&0\\
{{a_u}{e^{i{\alpha _u}}}}&{{d_{\rm{u}}}}&0\\
0&0&{{c_{\rm{u}}}}
\end{array}} \right)
\end{eqnarray}
or
\begin{eqnarray}
{M_u} = \left( {\begin{array}{*{20}{c}}
e_{u}{e^{ - 2i{\alpha _u}}}&{{a_u}{e^{ - i{\alpha _u}}}}&0\\
{{a_u}{e^{ - i{\alpha _u}}}}&{{d_{\rm{u}}}}&0\\
0&0&{{c_{\rm{u}}}}
\end{array}} \right),\;\nonumber
\end{eqnarray}
\begin{eqnarray}
{M_d} = \left( {\begin{array}{*{20}{c}}
0&{{a_d}}&0\\
{{a_d}}&{{d_{\rm{d}}}}&{{b_d}}\\
0&{{b_d}}&{{c_{\rm{d}}}}
\end{array}} \right)
\end{eqnarray}
where 
\begin{eqnarray}
{a_u} = \sqrt {\left( {{m_u} - {e_u}} \right)\left( {{m_c} + {e_u}} \right)},\;\nonumber\\
{d_u} = {m_u} - {m_c}-e_{u},\;\nonumber\\
{c_u} = {m_t},\;\nonumber\\
{{a_d}}= \sqrt {\frac{{{m_{\rm d}}{m_{\rm s}}{m_{\rm b}}}}{{{c_{\rm d}}}}},\;\nonumber\\
{b_d} = \sqrt {\frac{{({c_d} - {m_s})({c_d} + {m_d})({m_t} - {c_d})}}{{{c_d}}}} ,\;\nonumber\\
{c_{\rm d}} =  - {m_d} + {m_s} + {m_b}-d_d
\end{eqnarray}
involving three free parameters $e_{u}$, $d_{\rm d}$ and $\alpha_{u}$. Also the flavor mixing is independent of the quark mass $m_{\rm t}$. As a result, there are nine parameters involved in these and predictions may not be expected when $e_{u}\ne 0$. The resulting quark mixing matrix $V=O^{T}_{\rm u}PO_{\rm d}$, where $P=diag\lbrace e^{i\alpha_{u}},1,1\rbrace $ and
\begin{eqnarray}
{O_{\rm{u}}} = \left( {\begin{array}{*{20}{c}}
{\sqrt {\frac{{{m_c} + {e_u}}}{{({m_u} + {m_c})}}} }&{\sqrt {\frac{{{m_u} - {e_u}}}{{({m_u} + {m_c})}}} }&0\\
{\sqrt {\frac{{{m_u} - {e_u}}}{{({m_u} + {m_c})}}} }&{ - \sqrt {\frac{{{m_c} + {e_u}}}{{({m_u} + {m_c})}}} }&0\\
0&0&1
\end{array}} \right),
\end{eqnarray}
\begin{widetext}
\begin{eqnarray}
{O_d} = \left( {\begin{array}{*{20}{c}}
{\sqrt {\frac{{{m_s}{m_b}({c_d} + {m_d})}}{{{c_d}({m_b} + {m_d})({m_s} + {m_d})}}} }&{\sqrt {\frac{{{m_d}{m_b}({c_d} - {m_s})}}{{{c_d}({m_b} - {m_s})({m_d} + {m_s})}}} }&{\sqrt {\frac{{{m_d}{m_s}({m_b} - {c_d})}}{{{c_d}({m_b} - {m_s})({m_b} + {m_d})}}} }\\
{ - \sqrt {\frac{{{m_d}({c_d} + {m_d})}}{{({m_b} + {m_d})({m_s} + {m_d})}}} }&{\sqrt {\frac{{{m_s}({c_d} - {m_s})}}{{({m_b} - {m_s})({m_d} + {m_s})}}} }&{\sqrt {\frac{{{m_b}({m_b} - {c_d})}}{{({m_b} - {m_s})({m_b} + {m_d})}}} }\\
{\sqrt {\frac{{{m_d}({m_b} - {c_d})({c_d} - {m_s})}}{{{c_d}({m_b} + {m_d})({m_s} + {m_d})}}} }&{ - \sqrt {\frac{{{m_s}({c_d} + {m_d})({m_b} - {c_d})}}{{{c_d}({m_b} - {m_s})({m_d} + {m_s})}}} }&{\sqrt {\frac{{{m_b}({c_d} + {m_d})({c_d} - {m_s})}}{{{c_d}({m_b} - {m_s})({m_b} + {m_d})}}} }
\end{array}} \right).
\end{eqnarray}
\end{widetext}
Using $m_1<<m_2\sim d_{q}<<m_3$, one obtains
\begin{eqnarray}
{V_{us}} = \sqrt {\frac{{{m_d}}}{{{m_d} + {m_s}}}}  + \sqrt {\frac{{{m_u} - {e_u}}}{{{m_c}}}} {e^{ - i{\alpha _u}}},\;\nonumber\\
{V_{cb}} = \sqrt {\frac{{{m_d} - {m_s} + {d_d}}}{{{m_b}}}}, \;\nonumber\\
{V_{ub}} = \left( {\sqrt {\frac{{{m_d}{m_s}}}{{m_b^2}}}  + \sqrt {\frac{{{m_u} - {e_u}}}{{{m_c}}}} {e^{ - i{\alpha _u}}}} \right){V_{cb}}\nonumber \\
\end{eqnarray}
For $e_{u}=0$, the $V_{ub}$ equation indicates $m_{u}>3.8~MeV$ (i.e. deviation of over $5\sigma$ in $m_{u}$) is required for viability of $V_{ub}$ and $sin~2\beta$ with data. This is depicted in FIG. 8 and FIG. 9 respectively. Hence the corresponding texture five zero possibility appears to be ruled out by current precision measurements. This result is a consequence of the fact that $s_{13}$ is only generated at \textit{Next to Leading Order} for such texture structures. Note that $\beta\neq-Arg\lbrace V_{td}\rbrace$. 

The best-fit values for $V$ along with the various CP-angles and $J_{\rm{CP}}$ appear below, e.g.
\begin{eqnarray}
|V| = \left( {\begin{array}{*{20}{c}}
{0.974344}&{0.225031}&0.003557\\
{0.224880}&{0.973545}&{0.040464}\\
0.008982 &{0.039614}&{0.999174}
\end{array}} \right), \nonumber\\
\delta_{13}=76.96^{\circ},~~ 
J_{CP}=3.07\times 10^{-5},\nonumber\\
\alpha=80.98^{\circ},~~\beta=22.09^{\circ},~~\gamma=76.93^{\circ}.\end{eqnarray}
which correspond to
\begin{eqnarray}
{m_{\rm{u}}} = 4.47~{\rm{MeV}},~{m_{\rm{c}}} = 0.550~{\rm{GeV}},\nonumber\\
{m_{\rm{d}}} = 7.75~{\rm{MeV}},~ {m_{\rm{s}}} = 136~{\rm{MeV}},~{m_{\rm{b}}} = 2.86~{\rm{GeV}}.\;\nonumber\\
\end{eqnarray} 
Clearly, $m_{u}$, $m_{d}$ and $m_{s}$ must deviate over $5\sigma$ of current values. 

However, one obtains the following predictions for Case-VII when $e_{u}=-2m_{u}$,
\begin{eqnarray}
{V_{us}} = \sqrt {\frac{{{m_d}}}{{{m_d} + {m_s}}}}  + \sqrt {\frac{{3{m_u}}}{{{m_c}}}} {e^{ - i{\alpha _u}}},\;\nonumber\\
{V_{ub}} = \left( {\sqrt {\frac{{{m_d}{m_s}}}{{m_b^2}}}  + \sqrt {\frac{{3{m_u}}}{{{m_c}}}} {e^{ - i{\alpha _u}}}} \right){V_{cb}}
\end{eqnarray}
\begin{figure}
\includegraphics[scale=1.0]{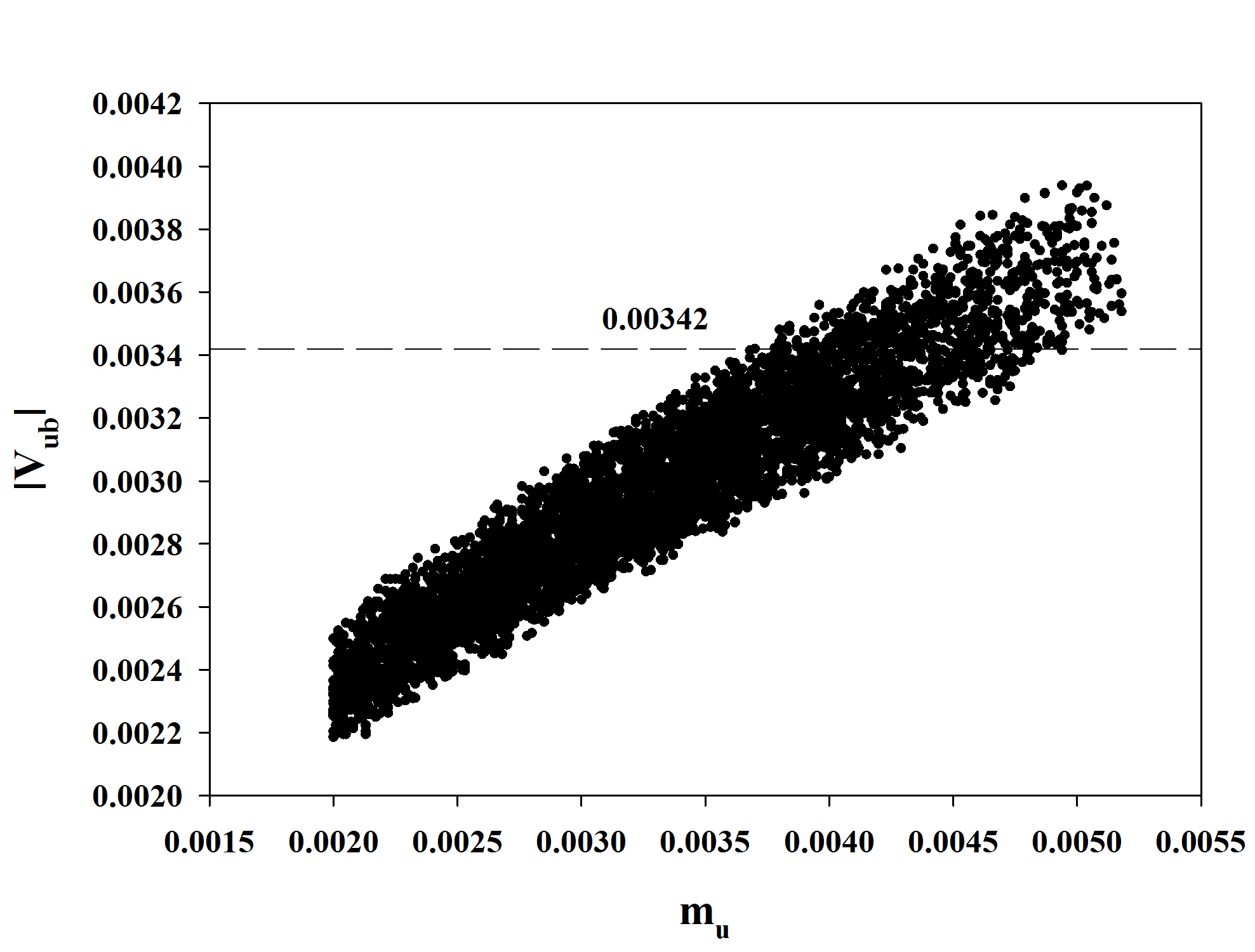}
\caption{$\mid V_{ub}\mid$ vs. $m_{u}$ for Case-VII with $e_{u}=0$}
\includegraphics[scale=1.0]{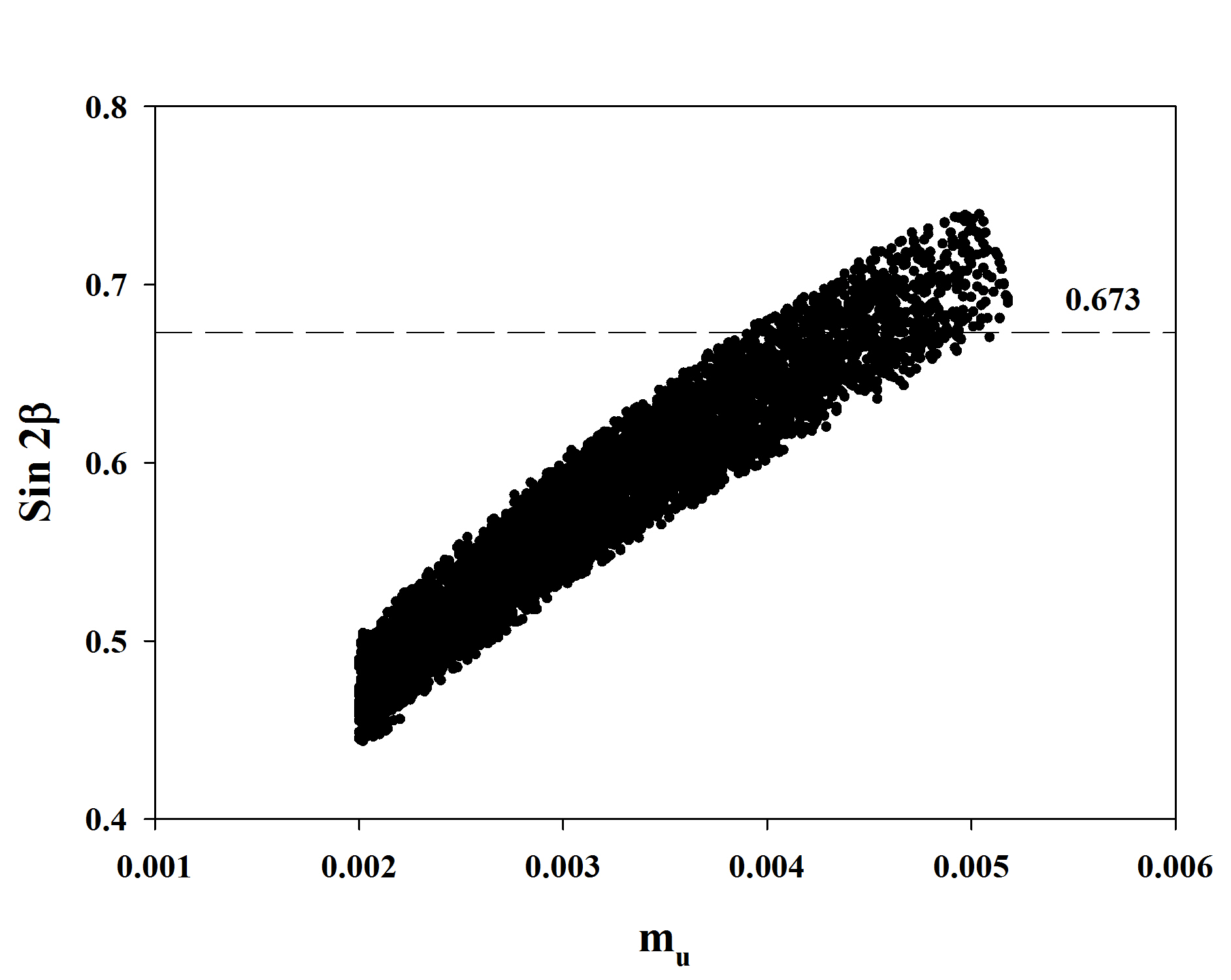}
\caption{$sin~2\beta$ vs. $m_{u}$ for Case-VII with $e_{u}=0$}
\end{figure}
The best-fit values for $V$ along with the various CP-angles and $J_{\rm{CP}}$ appear below, e.g.
\begin{eqnarray}
|V| = \left( {\begin{array}{*{20}{c}}
{0.974227}&{0.225539}&0.003552\\
{0.225380}&{0.973444}&{0.040131}\\
0.009188 &{0.039226}&{0.999188}
\end{array}} \right), \nonumber\\
\delta_{13}=81.31^{\circ},~~ 
J_{CP}=3.09\times 10^{-5},\nonumber\\
\alpha=76.85^{\circ},~~\beta=21.88^{\circ},~~\gamma=81.27^{\circ}.\;\nonumber\\
\end{eqnarray}
which correspond to
\begin{eqnarray}
\alpha_{u}=105.55^{\circ},\nonumber\\{m_{\rm{u}}} = 1.71~{\rm{MeV}},~~{m_{\rm{c}}} = 0.638~{\rm{GeV}},\nonumber\\
{m_{\rm{d}}} = 3.95~{\rm{MeV}},~~ {m_{\rm{s}}} = 68.4~{\rm{MeV}},~{m_{\rm{b}}} = 2.86~{\rm{GeV}}.\;\nonumber\\
\end{eqnarray} 
The corresponding quark mass matrices (in units of GeV) are
\begin{eqnarray}
\mid{M_{\rm{u}}}\mid = \left( {\begin{array}{*{20}{c}}
{ - 0.003420}&{0.057056}&{\bf{0}}\\
{0.057056}&{ - 0.632870}&{\bf{0}}\\
{\bf{0}}&{\bf{0}}&{172.1}
\end{array}} \right),\;\nonumber\\ 
{M_d} = \left( {\begin{array}{*{20}{c}}
0&{0.016450\;}&0\\
{0.016450\;}&{0.068987}&{0.112530}\\
0&{0.112530}&{2.855462}
\end{array}} \right). \;\nonumber\\ 
\end{eqnarray}
However, the texture four zero along with the corresponding texture five zero possibility obtained with textures interchanged in Case-VII are ruled out by $V_{ub}\simeq\sqrt{m_{u}/2m_{t}}$ requiring over 5$\sigma$ deviation in light quark masses for viability with current data.

\subsection{Case-VIII}
These texture four zero quark mass matrices with a minimal phase structure are expressed below
\begin{eqnarray}
{M_d} = \left( {\begin{array}{*{20}{c}}
0&{{a_d}{e^{ - i{\alpha _d}}}}&0\\
{{a_d}{e^{i{\alpha _d}}}}&{{d_{\rm{d}}}}&0\\
0&0&{{c_{\rm{d}}}}
\end{array}} \right)\;\nonumber\\ 
\end{eqnarray}
or
\begin{eqnarray}
{M_d} = \left( {\begin{array}{*{20}{c}}
0&{{a_d}{e^{ - i{\alpha _d}}}}&0\\
{{a_d}{e^{ - i{\alpha _d}}}}&{{d_{\rm{d}}}}&0\\
0&0&{{c_{\rm{d}}}}
\end{array}} \right),\;\nonumber
\end{eqnarray}
\begin{eqnarray}
{M_u} = \left( {\begin{array}{*{20}{c}}
e_{u}&{{a_u}}&0\\
{{a_u}}&{{d_{\rm{u}}}}&{{b_u}}\\
0&{{b_u}}&{{c_{\rm{u}}}}
\end{array}} \right)\;\nonumber\\ 
\end{eqnarray}
where 
\begin{eqnarray}
{a_d} = \sqrt {{m_d}{m_s}},~~
{d_d} = {m_d} - {m_s},~~
{c_d} = {m_b},\;\nonumber\\
{a_u} = \sqrt {\frac{{\left( {{m_u} + {e_u}} \right)\left( {{m_c} - {e_u}} \right)\left( {{m_t} - {e_u}} \right)}}{{\left( {{c_u} - {e_u}} \right)}}} ,\;\nonumber\\
{b_u} = \sqrt {\frac{{\left( {{m_u} + {c_u}} \right)\left( {{c_u} - {m_c}} \right)\left( {{m_t} - {c_u}} \right)}}{{\left( {{c_u} - {e_u}} \right)}}}  ,\;\nonumber\\
{c_{\rm u}} =  - {m_u} + {m_c} + {m_t}-d_u-e_u\;\nonumber\\
\end{eqnarray}
involving three free parameters $e_{u}$, $d_{\rm u}$ and $\alpha_{d}$. Also the flavor mixing is independent of the quark mass $m_{\rm b}$ and predictions may not be expected.
The resulting quark mixing matrix $V=O^{T}_{\rm u}PO_{\rm d}$, where $P=diag\lbrace e^{i\alpha_{d}},1,1\rbrace $ and
\begin{eqnarray}
{O_{\rm{d}}} = \left( {\begin{array}{*{20}{c}}
{\sqrt {\frac{{{m_s}}}{{({m_d} + {m_s})}}} }&{\sqrt {\frac{{{m_d}}}{{({m_d} + {m_s})}}} }&0\\
{\sqrt {\frac{{{m_d}}}{{({m_d} + {m_s})}}} }&{ - \sqrt {\frac{{{m_s}}}{{({m_d} + {m_s})}}} }&0\\
0&0&1
\end{array}} \right),\;\nonumber\\
\end{eqnarray}
\begin{widetext}
\begin{eqnarray}
{O_u} = \left( {\begin{array}{*{20}{c}}
{\sqrt {\frac{{({m_c} - {e_u})({m_t} - {e_u})({c_u} + {m_u})}}{{({c_u} - {e_u})({m_t} + {m_u})({m_c} + {m_u})}}} }&{\sqrt {\frac{{({m_u} + {e_u})({m_t} - {e_u})({c_u} - {m_c})}}{{({c_u} - {e_u})({m_t} - {m_c})({m_c} + {m_u})}}} }&{\sqrt {\frac{{({m_u} + {e_u})({m_c} - {e_u})({m_t} - {c_u})}}{{({c_u} - {e_u})({m_t} - {m_c})({m_t} + {m_u})}}} }\\
{ - \sqrt {\frac{{({m_u} + {e_u})({c_u} + {m_u})}}{{({m_t} + {m_u})({m_c} + {m_u})}}} }&{\sqrt {\frac{{({m_c} - {e_u})({c_u} - {m_c})}}{{({m_t} - {m_c})({m_c} + {m_u})}}} }&{\sqrt {\frac{{({m_t} - {e_u})({m_t} - {c_u})}}{{({m_t} - {m_c})({m_t} + {m_u})}}} }\\
{\sqrt {\frac{{({m_u} + {e_u})({m_t} - {c_u})({c_u} - {m_c})}}{{({c_u} - {e_u})({m_t} + {m_u})({m_c} + {m_u})}}} }&{ - \sqrt {\frac{{({m_c} - {e_u})({c_u} + {m_u})({m_t} - {c_u})}}{{({c_u} - {e_u})({m_t} - {m_c})({m_c} + {m_u})}}} }&{\sqrt {\frac{{({m_t} - {e_u})({c_u} + {m_u})({c_u} - {m_c})}}{{({c_u} - {e_u})({m_t} - {m_c})({m_t} + {m_u})}}} }
\end{array}} \right).
\end{eqnarray}
\end{widetext}
Note that $\beta\neq-Arg\lbrace V_{td}\rbrace$. However, the following interesting predictions are obtained for $d_{u}=3m_{c}/2$ and $e_{u}=m_{u}$,
\begin{eqnarray}
{V_{us}} = \sqrt {\frac{{{m_d}}}{{{m_d} + {m_s}}}}  + \sqrt {\frac{{2{m_u}}}{{{m_c}}}} {e^{ - i{\alpha _d}}},\;\nonumber\\
{V_{ub}} = \sqrt {\frac{{{m_u}}}{{{m_t}}}} {e^{ - i{\alpha _d}}},\;\nonumber\\
{V_{cb}} = \sqrt {\frac{{{m_c}}}{{2({m_t} - {m_c})}}}.
\end{eqnarray}
The best-fit values for $V$ along with the various CP-angles and $J_{\rm{CP}}$ appear below, e.g.
\begin{eqnarray}
|V| = \left( {\begin{array}{*{20}{c}}
{0.974337}&{0.225064}&0.003565\\
{0.224904}&{0.973536}&{0.040556}\\
0.009188 &{0.039662}&{0.999170}
\end{array}} \right), \nonumber\\
\delta_{13}=80.06^{\circ},~~ 
J_{CP}=3.12\times 10^{-5},\nonumber\\
\alpha=78.10^{\circ},~~\beta=21.88^{\circ},~~\gamma=80.02^{\circ}.
\end{eqnarray}
which correspond to
\begin{eqnarray}
\alpha_{d}=100.83^{\circ},~~{m_{\rm{d}}} = 3.92~{\rm{MeV}},~~{m_{\rm{s}}} = 73~{\rm{MeV}},\nonumber\\
{m_{\rm{u}}} = 2.17~{\rm{MeV}},~~ {m_{\rm{c}}} = 0.560~{\rm{GeV}},~{m_{\rm{t}}} = 172.1~{\rm{GeV}}.\;\nonumber\\
\end{eqnarray} 
The corresponding quark mass matrices (in units of GeV) are
\begin{eqnarray}
\mid{M_{\rm{u}}}\mid = \left( {\begin{array}{*{20}{c}}
{ - 0.003420}&{0.057056}&{\bf{0}}\\
{0.057056}&{ - 0.632870}&{\bf{0}}\\
{\bf{0}}&{\bf{0}}&{172.1}
\end{array}} \right),\;\nonumber\\ 
{M_d} = \left( {\begin{array}{*{20}{c}}
0&{0.016450\;}&0\\
{0.016450\;}&{0.068987}&{0.112530}\\
0&{0.112530}&{2.855462}
\end{array}} \right) \;\nonumber\\
\end{eqnarray}
which are in excellent agreement with current precision data.

\section{Results and Conclusion}
Given the plethora of quark masses and mixing data, a systematic analysis of hierarchical and symmetrically placed texture zero quark mass matrices has been carried out in the light of precision measurements of these parameters. A summary of the current status of texture specific quark mass matrices has also been presented to consolidate our understanding from a phenomenological point of view. 

In the absence of a compelling 'Top-Down' theory, we use the 'Bottom-Up' approach, and attempt to reproduce the mixing data using predictable structures involving a single non-trivial phase in $M_{q}$. The individual quark mass matrices are diagonalized through three successive quark rotations in 13, 12 and 23 quark planes and texture zeros are chosen in these mass matrices such that most of the three flavor mixing angles are generated predominantly from $M_{\rm u}$ or $M_{\rm d}$ at the leading order. This results in eight possible predictive structures, four each for texture five zeros and texture four zeros. 

Among these, the structures corresponding to Case-III, Case-V and Case-VI are observed to be completely ruled out by the current precision data. The remaining  five possibilities are not only consistent with the recent precision data upto 1$\sigma$, but also provide interesting relations among the flavor mixing angles and the quark mass ratios, which are presented in Table-III. In particular, some of the texture structures namely Case-I, Case-II, Case-VII and Case-VIII are observed to be highly sensitive to the light quark masses $m_u$, $m_d$ and $m_s$. It is also observed that a single non-trivial phase in these mass matrices is sufficient enough to account for the observed CP-violation in the quark sector. 

Particularly, for Case-I and Case-II, this phase is closely related and nearly equal to $\delta_{13}$ and one also obtains 
\begin{eqnarray}
\beta  =  - \arg \left( {1 - \frac{{{s_{13}}{c_{12}}}}{{{s_{12}}{s_{23}}}}{e^{ i{\delta _{{\rm{13}}}}}}} \right)=-Arg\lbrace V_{td}\rbrace.\;\nonumber\\
\end{eqnarray}
On the contrary, the relationship between the quark mass matrix phase and the CKM phase $\delta_{13}$ is not so trivial in the other three Cases. For all the viable texture structures listed in Table-III, we observe that the mass matrices are strongly hierarchical and that the Cabibbo angle is predominantly determined by the ratio $\sqrt{m_{d}/m_{s}}$ and corrections may also result from $\sqrt{m_{u}/m_{c}}$. It is also observed that $V_{cb}\cong \sqrt{m_d/m_b}$ or $V_{cb}\cong \sqrt{m_c/2m_t}$ along with $V_{ub}\simeq \sqrt{m_u/m_t}$ provide excellent agreements with the current mixing data. Therefore, we conclude that apart from the famous Fritzsch-like texture four zero quark mass matrices, which require at least two non-trivial phases in the quark mass matrices and only one prediction for Cabibbo angle, there may exist certain possibilities of texture five zero and texture four zero quark mass matrices, which are not only compatible with current precision data at level of 1$\sigma$ but also point towards compelling predictions among quark ratios and mixing angles that may be tested experimentally at the B-factories to extract further clues for fermion mass generation, flavor mixing and CP-violation.

In particular, a fine tuning of the free parameters in terms of the quark masses yields interesting predictions in certain texture possibilities as well as retain strongly hierarchical structures for quark mass matrices discussed in \cite{Verma:2015mgd}. In principle, it is concluded that in addition to Fritzsch-like texture four zeros, several viable structures for hierarchical quark mass matrices involving symmetric texture zeros may be compatible with current precision data, with certain structures being sensitive to light quark masses and all such structures \cite{Ludl:2015lta} may not yield interesting predictions as obtained in the Cases discussed above.

As a note, this phenomenological study has been carried out at the electo-weak $M_Z$ scale within the framework of the SM to test the predictive capabilities viz-a-viz the quark mass ratios and mixing angles for various texture zero possibilities and therefore the Renormalization Group Equation (RGE) based energy scale dependence of the quark masses and CKM parameters along with the stability of texture zeros has not been included herein. However, some recent analysis in this context \cite{PhysRevD.68.073008, Xing:2015sva} predict that the texture zeros in hierarchical quark mass matrices are quite stable to one-loop RGE effects. 
\begin{widetext}
\begin{center}
\begin{table}
\begin{tabular}{c c c c c c c}
\hline
Case & $M_{u}$ & $M_{d}$ & Predictions & $m_u$ & $m_d$ & $m_s$\\
\hline
I & $\left( {\begin{array}{*{20}{c}}
{\bf{0}}&{\bf{0}}& \times \\
{\bf{0}}& \times &{\bf{0}}\\
 \times &{\bf{0}}& \times 
\end{array}} \right)$ & $\left( {\begin{array}{*{20}{c}}
{\bf{0}}& \times &{\bf{0}}\\
 \times & \times & \times \\
{\bf{0}}& \times & \times 
\end{array}} \right)$ & $\begin{array}{*{20}{c}}
{\begin{array}{*{20}{c}}
{|{V_{us}}| = \sqrt {\frac{{{m_d}}}{{{m_s}}}} }\\
{|{V_{ub}}| = \sqrt {\frac{{{m_u}}}{{{m_t}}}} }\\
{{\delta _{13}} \simeq {\gamma _u}}
\end{array}}\\
{\beta  =  - Arg\{ {V_{td}}\} }
\end{array}$ & $2\sigma$ & $2\sigma$ & $1\sigma$\\
\hline
II & $\left( {\begin{array}{*{20}{c}}
 \times &{\bf{0}}&{\bf{0}}\\
{\bf{0}}& \times & \times \\
{\bf{0}}& \times & \times 
\end{array}} \right)$  & $\left( {\begin{array}{*{20}{c}}
{\bf{0}}& \times & \times \\
 \times & \times &{\bf{0}}\\
 \times &{\bf{0}}& \times 
\end{array}} \right)$ & $\begin{array}{*{20}{c}}
{|{V_{us}}| = \sqrt {\frac{{{m_d}}}{{{m_s}}}} }\\
\mid {V_{cb}}\mid  = \sqrt {\frac{{{m_c}}}{{2{m_t}}}} \\
{{\delta _{13}} = {\gamma _d}}\\
{\beta  =  - Arg\{ {V_{td}}\} }
\end{array}$ & $\times$ & $1\sigma$ & $1\sigma$ \\
\hline
IV & $\left( {\begin{array}{*{20}{c}}
{\bf{0}}&{\bf{0}}& \times \\
{\bf{0}}& \times & \times \\
 \times & \times & \times 
\end{array}} \right)$ & $\left( {\begin{array}{*{20}{c}}
{\bf{0}}& \times &{\bf{0}}\\
 \times & \times &{\bf{0}}\\
{\bf{0}}&{\bf{0}}& \times 
\end{array}} \right)$ &$\begin{array}{*{20}{c}}
{\left| {{V_{us}}} \right| = \left| {\sqrt {\frac{{{m_d}}}{{{m_s}}}}  - \sqrt {\frac{{{m_u}}}{{{m_c}}}} {e^{ - i{\alpha _d}}}} \right|}\\
{\left| {{V_{ub}}} \right| = \sqrt {\frac{{3{m_u}}}{{2{m_t}}}} }\\
{\;\left| {{V_{cb}}} \right| = \sqrt {\frac{{{m_c}}}{{2{m_t}}}} }
\end{array}$ & $1\sigma$ & $1\sigma$ & $1\sigma$\\
\hline
VII & $\left( {\begin{array}{*{20}{c}}
\times &\times&{\bf{0}}\\
\times&\times& {\bf{0}} \\
{\bf{0}}& {\bf{0}} & \times 
\end{array}} \right)$ & $\left( {\begin{array}{*{20}{c}}
{\bf{0}}&\times& {\bf{0}} \\
\times& \times & \times \\
{\bf{0}} & \times & \times 
\end{array}} \right)$ & $\begin{array}{*{20}{c}}
{\left| {{V_{us}}} \right| = \left| {\sqrt {\frac{{{m_d}}}{{{m_s}}}}  + \sqrt {\frac{{3{m_u}}}{{{m_c}}}} {e^{ - i{\alpha _u}}}} \right|}\\
{\frac{{{V_{ub}}}}{{{V_{cb}}}} = \sqrt {\frac{{{m_d}{m_s}}}{{m_b^2}}}  + \sqrt {\frac{{3{m_u}}}{{{m_c}}}} {e^{ - i{\alpha _u}}}\;}
\end{array}$ & $1\sigma$ & $1\sigma$ & $1\sigma$ \\
\hline
VIII & $\left( {\begin{array}{*{20}{c}}
\times&\times& {\bf{0}} \\
\times& \times & \times \\
{\bf{0}} & \times & \times 
\end{array}} \right)$ & $\left( {\begin{array}{*{20}{c}}
{\bf{0}} &\times&{\bf{0}}\\
\times&\times& {\bf{0}} \\
{\bf{0}}& {\bf{0}} & \times 
\end{array}} \right)$ & $\begin{array}{*{20}{c}}
{\left| {{V_{us}}} \right| = \left| {\sqrt {\frac{{{m_d}}}{{{m_s}}}}  + \sqrt {\frac{{2{m_u}}}{{{m_c}}}} {e^{ - i{\alpha _d}}}} \right|}\\
{\left| {{V_{ub}}} \right| = \sqrt {\frac{{{m_u}}}{{{m_t}}}} }\\
{\left| {{V_{cb}}} \right| = \sqrt {\frac{{{m_c}}}{{2{m_t}}}} }
\end{array}$ & $2\sigma$ & $2\sigma$ & $1\sigma$\\
\hline
\end{tabular}
\caption{Predictions from texture zeros and their dependence on light quark masses}
\end{table}
\end{center}
\end{widetext}

\begin{acknowledgments}
This work was supported in part by the Department of Science and Technology, India under SERB Research Grant No. SB/FTP/PS-140/2013.
\end{acknowledgments}
\bibliography{thebibliography}
\end{document}